\documentclass[runningheads]{llncs}
 
\usepackage[T1]{fontenc}
\usepackage{graphicx}
\usepackage{booktabs}
\usepackage[misc]{ifsym}
\newcommand{\corr}{(\Letter)}
\usepackage{mwe}
  
\usepackage[hidelinks]{hyperref}
\usepackage[utf8]{inputenc} 
\usepackage{amsmath,amssymb,amsfonts} 
\usepackage{booktabs}
\usepackage{algorithm}
\usepackage{algorithmic}

\usepackage[nolist,nohyperlinks]{acronym}  

\usepackage{tabularx}

\newcolumntype{R}{>{\raggedleft\arraybackslash}X}
\newcolumntype{L}{>{\raggedright\arraybackslash}X}
\newcolumntype{C}{>{\centering\arraybackslash}X}

\newcommand{\sd}[1]{\text{\tiny{}\textpm#1}}

\usepackage{subcaption}
\usepackage{wrapfig}
\usepackage{colortbl}
\usepackage{pifont}
\usepackage{graphicx} 
\usepackage{url}
\usepackage{makecell}
\usepackage{multirow}
\usepackage{pgf} 

\usepackage{colortbl}
\definecolor{grayrow}{gray}{0.9} 
\definecolor{myred}{HTML}{DC3912} 
\definecolor{myblue}{HTML}{3366CC} 
\definecolor{mygreen}{HTML}{109618} 

\newcommand{\xmark}{\ding{55}} 
\newcommand{\cmark}{\ding{51}}
 \usepackage{tikz}
\newcommand{\fullcircle}{\tikz{\filldraw[black] (0,0) circle (0.6ex);}}
\newcommand{\halfcircle}{%
    \tikz{
        \filldraw[black] (0,0) -- (0,0.6ex) arc[start angle=90, end angle=270, radius=0.6ex] -- cycle; 
        \draw[black] (0,0) circle (0.6ex); 
    }%
}
\newcommand{\emptycircle}{\tikz{\draw[black] (0,0) circle (0.6ex);}}

\begin{document}

\title{Graph Neural Networks for Jamming Source Localization}

\titlerunning{Graph Neural Networks for Jamming Source Localization}

\author{
    Dania Herzalla\textsuperscript{1}\textsuperscript{\dag}\corr \and 
    Willian T. Lunardi\textsuperscript{1}\textsuperscript{\dag} \and 
    Martin Andreoni\textsuperscript{1,2}
}
\renewcommand{\thefootnote}{\dag}
\footnotetext{These authors contributed equally.}

\institute{
    Technology Innovation Institute, Abu Dhabi, United Arab Emirates \\
    \email{\{dania.herzalla, willian.lunardi, martin.andreoni\}@tii.ae} 
    \and
    Computer Science Department, Khalifa University of Science and Technology, Abu Dhabi, United Arab Emirates
}

\authorrunning{D. Herzalla et al.}
\maketitle              

\begin{abstract}
Graph-based learning provides a powerful framework for modeling complex relational structures; however, its application within the domain of wireless security remains significantly underexplored. In this work, we introduce the first application of graph-based learning for jamming source localization, addressing the imminent threat of jamming attacks in wireless networks. Unlike geometric optimization techniques that struggle under environmental uncertainties and dense interference, we reformulate the localization as an inductive graph regression task. Our approach integrates structured node representations that encode local and global signal aggregation, ensuring spatial coherence and adaptive signal fusion. To enhance robustness, we incorporate an attention-based \ac{GNN} that adaptively refines neighborhood influence and introduces a confidence-guided estimation mechanism that dynamically balances learned predictions with domain-informed priors. We evaluate our approach under complex \ac{RF} environments with various sampling densities, network topologies, jammer characteristics, and signal propagation conditions, conducting comprehensive ablation studies on graph construction, feature selection, and pooling strategies. Results demonstrate that our novel graph-based learning framework significantly outperforms established localization baselines, particularly in challenging scenarios with sparse and obfuscated signal information. Our code is available at \textbf{\textit{\url{https://github.com/tiiuae/gnn-jamming-source-localization}}}.

\end{abstract}

\begin{acronym}
    \acro{GNN}{Graph Neural Network}
    \acro{GNNs}{Graph Neural Networks}
    \acro{FSPL}{Free Space Path Loss}
    \acro{LDPL}{Log Distance Path Loss}
    \acro{UAV}{Unmanned Aerial Vehicle}
    \acro{LOS}{Line-of-sight}
    \acro{NLOS}{Non-line-of-sight}
    \acro{ToA}{Time of Arrival} 
    \acro{TDoA}{Time Difference of Arrival} 
    \acro{DSP}{Digital Signal Processors}
    \acro{FPGA}{Field Programmable Gate Arrays} 
    \acro{COTS}{Commercial Off-The-Shelf} 
    \acro{RJSS}{Received Jamming Signal Strength}
    \acro{RSSI}{Received Signal Strength Indicator} 
    \acro{SNR}{Signal-to-noise ratio} 
    \acro{SINR}{Signal-to-interference-and-noise ratio} 
    \acro{PDR}{Packet Delivery Ratio}
    \acro{HPPP}{Homogeneous Poisson Point Process}
    \acro{FFT}{Fast Fourier Transform}
    \acro{CL}{Centroid Localization}
    \acro{WCL}{Weighted Centroid Localization}
    \acro{LSQ}{Least Squares}
    \acro{PL}{Path Loss}
    \acro{MLE}{Maximum Likelihood Estimation}
    \acro{MLAT}{Multilateration}
    \acro{RF}{Radio Frequency}
    \acro{RMSE}{Root Mean Squared Error}
    \acro{MSE}{Mean Squared Error}
    \acro{MAE}{Mean Absolute Error}
    \acro{KNN}{K-Nearest Neighbors}
    \acro{SDR}{Software Defined Radio}
    \acro{TPE}{Tree-structured Parzen Estimator}
    \acro{AoA}{Angle-of-Arrival}
    \acro{MLP}{Multilayer Perceptron}
    \acro{GCN}{Graph Convolutional Network}
    \acro{GAT}{Graph Attention Network}
    \acro{RNN}{Recurrent Neural Network}
    \acro{SGD}{Stochastic Gradient Descent}
    \acro{CAGE}{Confidence-guided Adaptive Global Estimation}
    \acro{PNA}{Principal Neighbourhood Aggregation}
\end{acronym}


\section{Introduction}
Graphs serve as a fundamental framework for representing complex relationships and interactions in real-world systems. Many highly successful machine learning applications are based on graph-based learning~\cite{velivckovic2023everything}. 
Although explored in numerous domains, the application of graph-based learning to wireless security remains underexplored. In particular, jamming source localization, a task essential for mitigating the threat of interference to the availability of wireless communication, presents a promising avenue for research. In this study, we investigate the application of a novel graph-based learning framework to perform an inductive graph-level regression task to predict the location of a jammer in complex \ac{RF} environments.

The ubiquitous and growing dependence on wireless networks for everyday connectivity and mission-critical operations introduces significant security vulnerabilities. As jamming attacks transmit intentional interference across communication channels, they lead to degradation and severing of wireless links~\cite{pelechrinis2009pdr}. The repercussions of such attacks are severe, leading to disruptions in essential services and operational hazards~\cite{wei2017survey}. Although countermeasures such as frequency hopping have been explored, their effectiveness is limited against advanced jammers such as reactive, follow-on, and barrage that adaptively pursue target frequencies creating dense interference across multiple channels~\cite{pirayesh2022jamattackssurvey}. Unlike avoidance techniques that passively adjust the network to evade interference, jammer localization provides a direct mitigation strategy, wherein network administrators can deploy countermeasures such as physical neutralization or geofencing to restore network reliability regardless of the attack strategy~\cite{wei2017survey}.

Classical jammer localization methods rely on geometric and optimization-based techniques to estimate the jammer's position. However, they degrade in real-world \ac{RF} environments due to noise and multipath effects~\cite{niu2020overview}. Their dependence on idealized propagation models limits adaptability to stochastic \ac{RF} dynamics. To overcome these limitations, we introduce a novel graph-based formulation of the jamming localization problem, leveraging attention-based \acp{GNN} to adaptively extract spatial and signal-related patterns from measurements. Additionally, we propose \ac{CAGE}, a confidence-guided estimation mechanism that dynamically balances GNN-based predictions with domain-informed priors, improving robustness in varying deployment conditions. Rather than treating localization as a geometric optimization problem, we redefine it as a graph regression task, where node features encode key RF and spatial characteristics, and the graph structure captures local and global signal dependencies. Our contributions are as follows:
\begin{itemize}
    \item We present the first application of \acp{GNN} to jamming source localization in wireless networks by reformulating the problem as an inductive graph regression task.
 
    \item We propose a graph-based learning framework with structured node representations for local and global signal aggregation. We also introduce a confidence-guided estimation mechanism to balance GNN predictions with domain-informed priors.

    \item We conduct comprehensive ablation studies on graph construction and model design, analyzing the impact of node connectivity, feature selection, pooling strategies, downsampling techniques, and graph augmentations on localization performance and model robustness.
    
    \item We benchmark against well-known localization methods in challenging environments, consistently outperforming established baselines, with emphasized improvements in scenarios characterized by sparse and obscured signal information.
\end{itemize}

The remainder of this paper is organized as follows. Section~\ref{sec:related_work} reviews related work, highlighting existing approaches to jamming source localization and their limitations. Section~\ref{sec:problem_definition} formally defines the problem, detailing the network configuration, jammer characteristics, and underlying assumptions. Section~\ref{sec:learning_framework} presents a learning framework that dynamically integrates data-driven representations with inductive priors. Section~\ref{sec:results} reports experimental results, demonstrating the effectiveness of our approach under various network conditions. Finally, we conclude by discussing key findings and future research directions.

\section{Related Work}\label{sec:related_work}

\subsection{Jamming Source Localization}\label{sec:jammer_localization}
Jamming source localization has been widely studied using range-free and range-based algorithms. Range-free algorithms perform localization using network topology-related properties, without relying on the physical characteristics of the incoming signals~\cite{hongbo2009vfil}. These methods are useful in infrastructure-limited environments, however, they suffer from degraded accuracy in sparse or unevenly distributed networks~\cite{wang2011weighted,wei2017survey}. In contrast, range-based methods utilize geometric optimization to estimate the distance to the source using the measurement of various physical properties such as \ac{RSSI}, \ac{ToA}, and \ac{AoA}~\cite{hongbo2009vfil}. While these methods generally achieve higher accuracy, obtaining measurements such as \ac{ToA} and \ac{AoA} rely on specialized and calibrated hardware and are susceptible to errors where multipath effects introduce significant biases~\cite{yan2023attnloc}. 

Many of the existing jammer localization approaches are validated under theoretical propagation models, such as free-space path loss, which fails to capture the complexities of real-world settings. Furthermore, many evaluations fail to include diversity in attack scenarios, sampling strategies, and long-range jamming effects~\cite{niu2020overviewloc}. To address these limitations, we evaluate our method on diverse network configurations, leveraging the \ac{LDPL}~\cite{etiabi2024metagraphloc} model to account for realistic signal propagation conditions (See Appendix~\ref{app:modeling}). 

As our method leverages spatial and signal information exclusively, we focus on methods that similarly harness this information for localization. We benchmark our approach against established range-free and range-based localization techniques: \ac{WCL}~\cite{wang2011weighted}, \ac{LSQ}~\cite{wei2017survey}, \ac{PL}~\cite{nardin2023apbm}, \ac{MLE}~\cite{nardin2023apbm}, and \ac{MLAT}~\cite{yang2016mlat}. These baselines serve as reference points to assess the robustness and adaptability of our graph-based framework.

\subsection{Graph Neural Networks for Localization}\label{sec:gnn_localization}

\acp{GNN} have demonstrated strong capabilities in regression tasks by capturing spatial dependencies within graph-structured data, such as in molecular property prediction and material science~\cite{merchant2023scaling}. Beyond these domains, \acp{GNN} have also been applied to pose regression problems, including camera pose estimation~\cite{turkoglu2021viscamreloc} and human pose tracking~\cite{yang2021learning}, where they refine node and edge representations to improve motion prediction and spatial consistency. Their effectiveness in localization tasks has also gained attention, particularly in wireless and sensor network applications. \acp{GNN} have been applied to RF-based localization, including WiFi fingerprinting-based indoor localization~\cite{lezama2023gnnindoor}, where they leverage \ac{RSSI} signals to construct graphs and improve positioning accuracy. More recent works extend \acp{GNN} for network localization beyond Wi-Fi, addressing challenges such as dynamic network topologies and mobility-induced signal variation~\cite{etiabi2024metagraphloc}.

While these advancements highlight the growing potential of \acp{GNN} for network localization, their application to jamming source localization remains unexplored. This gap presents an opportunity to adapt \acp{GNN} to interference localization, addressing unique challenges under adversarial conditions.

\section{Problem Definition}
\label{sec:problem_definition}
The problem addressed in this study is the localization of a wireless jammer, an adversarial interference source that disrupts communications by emitting intentional interference signals to degrade legitimate wireless communication.  

\subsubsection{Network Configuration}
The network consists of \( N \) devices, either static or dynamic, deployed in a \( D \)-dimensional space, where \( \mathcal{A} \subset \mathbb{R}^D \) defines the geographic area of interest and \( N \geq 1 \). Each device \( i \) records signal measurements over time, forming a set of samples: 
\[
    \mathcal{S}_i = \{s_i^1, s_i^2, \dots, s_i^T\}, \quad s_i^t = (\mathbf{x}_i^t, \eta_i^t)
\]
where \( s_i^t \) denotes a measurement taken at position \( \mathbf{x}_i^t \) with an associated noise floor value \( \eta_i^t \) in dBm at time \( t \). Note that \( \eta_i^t \), referred to as the \textit{noise floor}, is commonly treated as jamming signal strength, where here it represents the combined effect of jammer interference and baseline environmental noise floor. Signal attenuation and propagation effects are modeled using the empirical \ac{LDPL} model to simulate \ac{NLOS} conditions; refer to Appendix~\ref{app:modeling} for details.

This formulation accommodates both \textit{static} and \textit{dynamic} sampling scenarios. In the static case, devices remain at fixed locations but are likely distributed over space. In contrast, in the dynamic case, devices move through space, collecting measurements at different positions over time.

\subsubsection{Jammer Characteristics}
The jammer is located at an unknown position \( \mathbf{x}_j \in \mathcal{A} \) and emits interference signals that elevate the noise floor of nearby devices. The interference strength varies spatially due to distance attenuation and propagation effects such as shadowing and multipath fading. 
The sampled region, denoted as \( \mathcal{R} \subseteq \mathcal{A} \), is the subset of \( \mathcal{A} \) where devices collect noise floor measurements. Although the jammer's position satisfies \( \mathbf{x}_j \in \mathcal{A} \), it does not necessarily hold that \( \mathbf{x}_j \in \mathcal{R} \), i.e., \( \mathbf{x}_j \in \mathcal{A} \setminus \mathcal{R} \) is possible. We evaluate localization methods in scenarios where the jammer affects areas beyond the sampled region, testing their ability to infer positions outside direct measurement zones.

\subsubsection{Objective}
Given the set of samples collected by \( N \) devices, denoted as \( \mathcal{S} = \{\mathcal{S}_1, \mathcal{S}_2, \dots, \mathcal{S}_N\} \), the goal is to infer \( \mathbf{x}_j \) based on the spatial distribution of measured noise floor levels. This problem is inherently challenging due to environmental noise and uncertainty affecting signal propagation, complex spatial correlations between noise floor levels and jammer interference, and the need to generalize beyond observed regions where no direct measurements are available.

\section{Graph-Structured Learning for Jammer Localization}\label{sec:learning_framework}

Given the set of collected measurements \( \mathcal{S} = \{\mathcal{S}_1, \mathcal{S}_2, \dots, \mathcal{S}_N\} \), we represent the sampled signal space as a graph \( G = (V, E) \), where nodes in \( V \) correspond to individual measurement instances \( s_i^t \in \mathcal{S}_i \), and edges \( E \) define spatial relationships between them. Edges are established using \ac{KNN}, where for each node \( v_i \), an edge set is constructed as \( E = \{(v_i, v_j) \mid v_j \in \mathcal{N}_k(v_i)\} \), where \( \mathcal{N}_k(v_i) \) denotes the set of \( k \) nearest neighbors of \( v_i \) in the Euclidean space. To enforce the spatial attenuation principle of jamming signals, each edge \( (v_i, v_j) \) is assigned a weight \( w_{ij} \) that decays exponentially with the Euclidean distance as:
\begin{equation}\label{eq:edge_weight}
    w_{ij} = \frac{e^{-d_{ij}} \left(e - e^{d_{ij}}\right)}{e - 1},
\end{equation}
where \( d_{ij} = \frac{\|\mathbf{x}_i - \mathbf{x}_j\|}{d_{\max}} \) is the normalized Euclidean distance between nodes \( v_i \) and \( v_j \), with \( d_{\max} = \max_{(v_i, v_j) \in E} \|\mathbf{x}_i - \mathbf{x}_j\| \).

Since message passing in the graph relies on node and edge attributes, we define a structured representation that incorporates both spatial and signal characteristics. Each node \( i \in V \) is assigned a feature vector:
\begin{equation}\label{eq:feature_vector}
    X_i = (\tilde{\eta}_i, \mathbf{x}_i^{\text{sph}}, \mathbf{x}_i^{\text{cart}}, \mathcal{F}_i^{\text{local}}).
\end{equation}
where \( \tilde{\eta}_i \) is the normalized noise floor value and \( \mathbf{x}_i^{\text{sph}} \) represents the normalized angular representation:
\begin{equation}\label{eq:five_tuple_rep}
    \mathbf{x}_i^{\text{sph}} = (r_i, \sin\theta_i, \cos\theta_i, \sin\phi_i, \cos\phi_i),
\end{equation} 
where \( r_i \) is the radial distance from the origin, and \( (\theta_i, \phi_i) \) are the azimuth and elevation angles of the measured position. We additionally incorporate the normalized cartesian coordinate representation \( \mathbf{x}_i^{\text{cart}} = (\Tilde{x}_i, \Tilde{y}_i, \Tilde{z}_i) \) to maintain direct Euclidean relationships between nodes. 

To encode spatial correlations and local noise floor variations within the graph, we define \( \mathcal{F}_i^{\text{local}} \), which characterizes the local noise floor distribution within each node's neighborhood:
\[
    \mathcal{F}_i^{\text{local}} = \{ \text{median}(\eta_{\mathcal{N}_k(i)}), \max(\eta_{\mathcal{N}_k(i)}), \Delta\eta_i, \mathbf{x}_i^{\text{wcent}}, d_i^{\text{wcent}} \},
\]
where \( \text{median}(\eta_{\mathcal{N}_k(i)}) \) and \( \max(\eta_{\mathcal{N}_k(i)}) \) provide local statistical summaries of the noise floor levels, \( \Delta\eta_i \) represents the deviation of the node's noise level from the mean noise level within its neighborhood, and \( \mathbf{x}_i^{\text{wcent}} \) and \( d_i^{\text{wcent}} \) correspond to the \textit{local weighted centroid} and its distance from the node \( v_i \), respectively:
\[
    \mathbf{x}_i^{\text{wcent}} = \frac{\sum_{j \in \mathcal{N}_k(i)} \eta_j' \mathbf{x}_j}{\sum_{j \in \mathcal{N}_k(i)} \eta_j'}, \quad 
    d_i^{\text{wcent}} = \|\mathbf{x}_i - \mathbf{x}_i^{\text{wcent}}\|, \quad 
    \Delta\eta_i = \eta_i - \frac{1}{|\mathcal{N}_k(i)|} \sum_{j \in \mathcal{N}_k(i)} \eta_j,
\]
where $\eta_j' = 10^{\eta_j / 10}$ represents the noise floor level converted to linear scale.

For dynamic scenarios where devices move while collecting measurements, the graph representation is extended to incorporate temporal dependencies. Each node \( i \in V \) retains its spatial attributes while additionally capturing motion through two features: the \textit{direction vector} and the \textit{temporal signal variation}. These measurements, computed between consecutive positions, are given, respectively, as:
\[
    \mathbf{d}_i = \mathbf{x}_i^{t+1} - \mathbf{x}_i^t, \quad \Delta \eta_{i}^\text{temp} = \eta_i^{t+1} - \eta_i^t.
\]
The final node feature vector for dynamic scenarios is given by:
\begin{equation}
    X_i = (\tilde{\eta}_i, \mathbf{x}_i^{\text{sph}}, \mathbf{x}_i^{\text{cart}}, \mathcal{F}_i^{\text{local}}, \mathbf{d}_i, \Delta \eta_{i}^\text{temp}).
\end{equation}

\subsection{Learning Spatial Relations with Attention-Based Graphs}

Graph-based learning enables the model to capture structured dependencies in the jammer interference field, allowing for improved generalization across spatial regions. However, effective learning in this setting requires handling two key challenges: (1) signal measurements are inherently noisy due to environmental conditions, leading to unreliable observations, and (2) the importance of neighboring nodes varies depending on both their spatial proximity and their reliability in capturing interference effects. Traditional GNN architectures, such as \acp{GCN}~\cite{kipf2017gcn}, perform uniform neighborhood aggregation, which limits their ability to weigh the informativeness of spatially adjacent nodes. Our approach follows \ac{GAT}~\cite{veličković2018gat}, incorporating an adaptive weighting mechanism that dynamically refines neighborhood influence, ensuring that message passing prioritizes nodes with reliable signal information while attenuating contributions from potentially misleading observations.

Let \( X \in \mathbb{R}^{|V| \times F} \) represent the matrix of raw node features, where each node \( v_i \) has an initial feature vector \( h_i^{(0)} = X_{i} \in \mathbb{R}^F \). Before computing attention, node features are first transformed at each layer \( l \) using a learnable weight matrix \( \mathbf{W}^{(l)} \). The updated node representation at layer \( l+1 \) is computed through the aggregation function denoted as:
\begin{equation}\label{eq:attention_aggregation}
    h_i^{(l+1)} = \text{ReLU}\left(\sum_{j \in \mathcal{N}_i} \alpha_{ij}^{(l)} \mathbf{W}^{(l)} h_j^{(l)} \right).
\end{equation}
The attention coefficients \( \alpha_{ij}^{(l)} \), weighted by \( w_{ij} \), determine the relative importance of each neighboring node \( v_j \) to node \( v_i \) and are computed as:
\begin{equation}
    \alpha_{ij}^{(l)} = \frac{w_{ij} \exp \left( \text{LeakyReLU} \left(\mathbf{a}^{T} [\mathbf{W}^{(l)} h_i^{(l)} \| \mathbf{W}^{(l)} h_j^{(l)}] \right) \right)}
    {\sum_{k \in \mathcal{N}(i)} w_{ik} \exp \left( \text{LeakyReLU} \left(\mathbf{a}^{T} [\mathbf{W}^{(l)} h_i^{(l)} \| \mathbf{W}^{(l)} h_k^{(l)}] \right) \right)},
\end{equation}
where \( \mathbf{a} \in \mathbb{R}^{2F'} \) is a learnable attention weight vector, with \( 2F' \) corresponding to the concatenation of intermediate node embeddings of dimension \( F' \). Here, \( (\cdot)^T \) denotes transposition, \( \| \) represents vector concatenation, and \( w_{ij} \) is given by Equation~\eqref{eq:edge_weight}. In our implementation, we utilize multi-head attention as originally proposed in \cite{veličković2018gat}, where multiple independent attention mechanisms operate in parallel. The resulting node embeddings are concatenated across attention heads, except in the final layer, where they are averaged. 

\subsection{Supernode-Guided Adaptive Estimation}

Prior work~\cite{wang2011weighted} establishes that \ac{WCL}, a simple localization method based on weighted averaging of anchor positions, achieves low localization error when node density around jammer's position is high and placement is radially symmetric, with error decreasing as the number of nodes increases. Building on this, we expand the graph definition to incorporate \ac{WCL} as a domain-informed prior, leveraging it under dense and symmetrical sampling conditions while introducing an adaptive confidence weighting mechanism for effective integration within the learning framework.

\subsubsection{Global Context Encoding with Domain-Guided Priors}
We extend the graph representation \( G = (V, E) \) by introducing a \textit{supernode} \( v_s \), which encodes a structured global prior by representing the weighted centroid of the measurement space based on noise floor levels. The augmented graph is defined as:
\begin{equation}
    G' = (V', E'), \quad \text{where} \quad V' = V \cup \{ v_s \}, \quad E' = E \cup \{(v_i \to v_s) \mid v_i \in V \}.
\end{equation}
Each node \( v_i \in V \) is connected to the supernode via a directed edge \( (v_i \to v_s) \) with weight \( w_{is} \) given by Equation~\eqref{eq:edge_weight}. This connectivity structure ensures that the supernode functions as a global aggregator, primarily influencing the computation of a confidence weight \( \alpha \) (later defined in Equation~\eqref{eq:final_prediction}) while remaining decoupled from the GNN regression process, thereby preventing direct bias from the WCL prior. The spatial position and noise level of the supernode are defined as:
\begin{equation}\label{eq:supernode_position}
    \mathbf{x}_{\text{super}} = \frac{\sum_{i \in V} w_i \mathbf{x}_i}{\sum_{i \in V} w_i}, \quad
    \eta_{\text{super}} = \frac{\sum_{i \in V} w_i \eta_i}{\sum_{i \in V} w_i}, \quad
    \text{where } w_i = \frac{\eta'_i}{\sum_{j \in V} \eta'_j}.
\end{equation}
Here, \( w_i \) represents the normalized weight assigned to each node \( v_i \). Since the feature vector for each node, as defined in Equation~\ref{eq:feature_vector}, is a function of position, noise, and neighboring nodes, we also expand the feature representation for \( v_s \).

\begin{figure}[!t]
    \centering
    \includegraphics[width=1.0\linewidth]{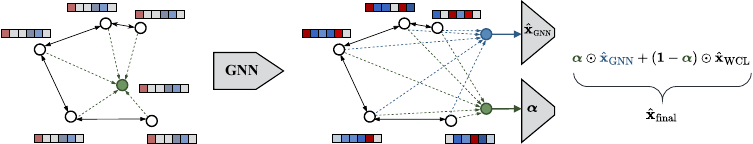}
    \caption{Overview of the proposed jammer localization framework. A graph is constructed where nodes represent spatial and signal instances, and edges capture spatial relationships. The encoder processes the graph to learn spatial correlations. The final jammer position is estimated through an adaptive combination of the GNN prediction and WCL prior, controlled by a learned confidence weight.}
    \label{fig:method}
\end{figure}

\subsubsection{Confidence-Guided Adaptive Position Estimation}
The estimated jammer position is computed as a five-tuple normalized angular representation combining the GNN-based prediction with the domain-informed WCL prior. The GNN-based position estimate is obtained by applying a linear transformation to a pooled representation of the node embeddings:
\begin{equation}
    \hat{\mathbf{x}}_{\text{GNN}} = \mathbf{W}_{\text{GNN}} h_{\text{graph}} + \mathbf{b}_{\text{GNN}},
\end{equation}
where \( \hat{\mathbf{x}}_{\text{GNN}} = (\hat{r}, \hat{s}_{\theta}, \hat{c}_{\theta}, \hat{s}_{\phi}, \hat{c}_{\phi}) \) represents the predicted position, with \( \mathbf{W}_{\text{GNN}} \in \mathbb{R}^{5 \times F'} \) and \( \mathbf{b}_{\text{GNN}} \in \mathbb{R}^{5} \) as learnable parameters. The graph representation \( h_{\text{graph}} \) of $G'$ is computed using an element-wise max pooling operation over all node embeddings, excluding the supernode:
\begin{equation}
    h_{\text{graph}} = \max_{v_i \in V} h_i^{(L)},
\end{equation}
where \( h_i^{(L)} \) is the final embedding of node \( v_i \) after \( L \) layers of attention-based aggregation given in Equation~\eqref{eq:attention_aggregation}. To determine the confidence weights \( \boldsymbol{\alpha} = (\alpha_1, \alpha_2, \alpha_3, \alpha_4, \alpha_5) \), the supernode representation \( h_{\text{super}} \) is passed through a linear transformation followed by a sigmoid activation:
\begin{equation}
    \boldsymbol{\alpha} = \sigma(\mathbf{W}_{\alpha} h_{\text{super}} + \mathbf{b}_{\alpha}),
\end{equation}
where \( \boldsymbol{\alpha} \in \mathbb{R}^5 \) is a five-dimensional confidence vector, with each \(\alpha_d\) corresponding to one of the five output components in the normalized angular representation. The parameters \( \mathbf{W}_{\alpha} \in \mathbb{R}^{5 \times F'} \) and \( \mathbf{b}_{\alpha} \in \mathbb{R}^{5} \) are learnable, and sigmoid \( \sigma(\cdot) \) ensures that \( 0 < \alpha_d < 1 \). Finally, the predicted jammer position is computed as:
\begin{equation}\label{eq:final_prediction}
    \hat{\mathbf{x}}_{\text{final}} = \boldsymbol{\alpha} \odot \hat{\mathbf{x}}_{\text{GNN}} + ( \mathbf{1} - \boldsymbol{\alpha} ) \odot \hat{\mathbf{x}}_{\text{WCL}},
\end{equation}
where \( \odot \) denotes element-wise multiplication, and \( \hat{\mathbf{x}}_{\text{WCL}} \) is the five-tuple normalized angular representation of the WCL estimate.

This formulation allows the model to adaptively balance the reliance on the GNN-based prediction and the structured WCL prior, ensuring robustness across varying sampling densities and spatial distributions. The corresponding graph-based formulation and adaptive position estimation process are illustrated in Figure~\ref{fig:method}, showing the transformation from raw signal measurements to the final position estimate through graph construction, GNN encoding, and confidence-weighted integration of the WCL prior.

\subsection{Training Strategy and Loss Function}

To enable adaptive estimation, we define a loss function that minimizes localization error by optimizing the weighted combination of the GNN-based estimate and the WCL prior. Given a batch of training instances \( \mathcal{B} \), the loss function for adaptive estimation is formulated as:
\begin{equation}\label{eq:loss_adapt}
    \mathcal{L}_{\text{Adapt}} = \frac{1}{|\mathcal{B}|} \sum_{m \in \mathcal{B}} \Big\| \hat{\mathbf{x}}_j^{(m)} - \big( \boldsymbol{\alpha}^{(m)} \odot \hat{\mathbf{x}}_{\text{GNN}}^{(m)} + ( \mathbf{1} - \boldsymbol{\alpha}^{(m)} ) \odot \hat{\mathbf{x}}_{\text{WCL}}^{(m)} \big) \Big\|^2,
\end{equation}
where \( \hat{\mathbf{x}}_j^{(m)} \) is the ground truth jammer position in the normalized angular representation, \( \hat{\mathbf{x}}_{\text{GNN}}^{(m)} \) is the predicted position from the GNN model, and \( \hat{\mathbf{x}}_{\text{WCL}}^{(m)} \) is the WCL-based position estimate. The confidence vector \( \boldsymbol{\alpha}^{(m)} \in \mathbb{R}^5 \) is learned from the supernode representation and dynamically balances the contribution of the two estimations.

\subsubsection{Encouraging Independent Learning of the GNN Regressor}
While the confidence mechanism allows optimal weighting of WCL and GNN estimates, an inherent risk of optimizing Equation~\eqref{eq:loss_adapt} alone is that the GNN regressor might learn primarily as a residual corrector for WCL rather than as an independent position estimator. To prevent this, we introduce an additional loss term that enforces direct learning of the jammer’s position by the GNN. This leads to the joint loss formulation:
\begin{equation}\label{eq:loss_cage}
    \mathcal{L}_{\text{CAGE}} = \frac{1}{2} \left( \mathcal{L}_{\text{GNN}} + \mathcal{L}_{\text{Adapt}} \right)
    + \lambda \sum_{m \in \mathcal{B}} (1 - \boldsymbol{\alpha}^{(m)})^2,
\end{equation}
where
\begin{equation}\label{eq:loss_gnn}
    \mathcal{L}_{\text{GNN}} = \frac{1}{|\mathcal{B}|} \sum_{m \in \mathcal{B}}  
    \Big\| \hat{\mathbf{x}}_j^{(m)} - \hat{\mathbf{x}}_{\text{GNN}}^{(m)} \Big\|^2.
\end{equation}
Here, \(\mathcal{L}_{\text{GNN}}\) ensures that the GNN independently learns to predict the jammer’s position without being influenced by WCL, while \( \mathcal{L}_{\text{Adapt}} \) (as defined in Equation~\eqref{eq:loss_adapt}) optimizes the weighted combination of GNN and WCL estimates, ensuring that the model learns to assign appropriate confidence to each based on spatial conditions. The final term in Equation~\eqref{eq:loss_cage} penalizes deviations of \( \boldsymbol{\alpha}^{(m)} \) from 1, reducing over-reliance on WCL. For simplicity, we set $\lambda = 0$ in our experiments while retaining this term for flexibility.

Note that in the experimental evaluation, we refer to our proposed method as \ac{CAGE}. For clarity in comparisons, \ac{MLP}, \ac{GCN}, \ac{PNA} and \ac{GAT} operate on graph \( G \) and are trained with Equation~\eqref{eq:loss_gnn} with the final estimate given by $\hat{\mathbf{x}}_{\text{GNN}}$, while CAGE is evaluated on the augmented graph \( G' \) and the adaptive confidence-weighted estimation, trained with Equation~\eqref{eq:loss_cage}.

\section{Experimental Evaluation}\label{sec:results}

We evaluate \ac{CAGE} across static and dynamic environments under the LDPL model. In the static setting (Section~\ref{sec:static}), fixed nodes with varying positions and densities are considered, with the jammer positioned either inside or outside the sampled region \( \mathcal{R} \) to study spatial effects.
In the dynamic setting (Section~\ref{sec:dynamic}), a device moves in 3D space, approaching and encircling the jammer to assess localization accuracy across angles and distances.
Ablation studies in Section~\ref{sec:components} and Appendix~\ref{app:ablations} evaluate the impact of node features, edge construction, graph augmentations, global pooling, and downsampling techniques.
Details on data generation, node spatial arrangements, jammer characteristics, and signal propagation environments varied in our experiments are provided in Appendix~\ref{app:modeling}, and an analysis of confidence weighting is presented in Section~\ref{sec:components}.

We compare \ac{CAGE} against classical methods (\ac{WCL}~\cite{wei2017survey}, \ac{LSQ}~\cite{wei2017survey}, \ac{PL}~\cite{nardin2023apbm}, \ac{MLAT}~\cite{yang2016mlat}, \ac{MLE}~\cite{nardin2023apbm}, MLP) and graph-based learning methods (\ac{GCN}~\cite{kipf2017gcn}, \ac{PNA}~\cite{corso2020pna}, \ac{GAT}~\cite{veličković2018gat}). As previously described, while MLP, \ac{GCN}, \ac{PNA}, and \ac{GAT} operate on \( G \), \ac{CAGE} leverages augmented graph \( G' \) with the supernode, incorporating graph attention mechanisms alongside confidence-weighted estimation for adaptive localization. Models are trained using AdamW with a one-cycle cosine annealing scheduler. Hyperparameter tuning details, including model architectures and optimizer settings, are provided in Appendix~\ref{app:hyperparams_modelarch}. Appendix~\ref{app:downsampling} describes the downsampling techniques applied in the dynamic experiments. All experiments are conducted over three independent trials with different random seeds to ensure robust evaluation, reporting the mean and standard deviation of \ac{MAE} and \ac{RMSE} as the primary evaluation metrics.  

\begin{table*}[t!]
    \centering
    \scriptsize
    \setlength{\tabcolsep}{0pt}
    \renewcommand{\arraystretch}{1.0}
    \caption{\ac{RMSE} in jammer localization for static scenarios, averaged over three trials with different seeds. Results are split by sampling geometry. MAE results are provided in Appendix~\ref{app:further_results} Table~\ref{tab:mae_fspl_ldpl_dataset_exp}.}
    \label{tab:rmse_fspl_ldpl_dataset_exp} 
    \begin{tabularx}{\textwidth}{@{} p{0.25cm} p{1cm} *{5}{C} *{5}{C} p{0.75cm}@{}}
    \toprule
    & \multirow{2}{*}{\textbf{Method}} & \multicolumn{5}{c}{\textbf{Jammer within ($\mathbf{x}_j \in \mathcal{R}$)}}  
    & \multicolumn{5}{c}{\textbf{Jammer outside ($\mathbf{x}_j \in \mathcal{A} \setminus \mathcal{R} $)}} 
    & \multirow{2}{*}{\textbf{Mean}} \\
    \cmidrule(lr){3-7} \cmidrule(lr){8-12}
    & & \textbf{C}  & \textbf{T} & \textbf{R} & \textbf{RD} & \textbf{Mean} 
      & \textbf{C} & \textbf{T} & \textbf{R} & \textbf{RD} & \textbf{Mean} & \\
    \midrule
    \multirow{10}{*}{\rotatebox[origin=c]{90}{RMSE}} 
    & \, WCL & 53.6 & 65.6 & 39.7 & 54.1 & 53.3 & 201.6 & 241.1 & 254.6 & 234.0 & 232.8 & 143.1 \\ 
    & \, PL & 159.5 & 115.0 & 121.3 & 114.6 & 127.6 & 357.3 & 314.4 & 365.2 & 336.2 & 343.3 & 235.5 \\ 
    & \, MLE & 123.2 & 112.5 & 116.5 & 362.3 & 178.6 & 295.8 & 302.9 & 318.3 & 1002.7 & 479.9 & 329.3 \\ 
    & \, MLAT & 158.1 & 125.5 & 110.4 & 98.2 & 123.1 & 346.8 & 329.7 & 353.4 & 309.3 & 334.8 & 229.0 \\ 
    & \, LSQ & 299.0 & 268.9 & 146.7 & 487.5 & 300.5 & 495.6 & 440.5 & 568.8 & 713.9 & 554.7 & 427.6 \\ 
    \cmidrule(lr){2-13}
    & \, MLP & 54.1 & 46.1 & 34.7 & 42.6 & 44.4 & 95.7 & 120.3 & 120.6 & 125.0 & 115.4 & 79.9 \\  
    & \, GCN & 51.6 & 44.2 & 36.1 & 49.3 & 45.3 & 91.5 & 115.1 & 117.7 & 124.3 & 112.2 & 78.8 \\   
    & \, PNA & 50.8 & 41.3 & 30.6 & 38.8 & 40.4 & 91.1 & 115.0 & 113.9 & 119.7 & 109.9 & 75.2 \\
    & \, GAT & 49.7 & 41.1 & 30.1 & 39.2 & 40.0 & 89.6 & 113.7 & 114.3 & 117.6 & 108.8 & 74.4 \\  
    & \, CAGE & \textbf{42.8} & \textbf{36.5} & \textbf{27.9} & \textbf{35.7} & \textbf{35.7} & \textbf{77.2} & \textbf{101.4} & \textbf{104.1} & \textbf{107.3} & \textbf{97.5} & \textbf{66.6} \\  
    \bottomrule
    \end{tabularx}
\end{table*}

\subsection{Static Evaluation for Jamming Localization}\label{sec:static}

To analyze the impact of node arrangements and coverage on localization accuracy, we evaluate performance in a static setting where devices remain fixed at predefined locations. Each instance consists of nodes randomly placed within the geographic area \(\mathcal{A} = \{ (x, y) \in \mathbb{R}^2 \mid 0 \leq x, y \leq 1500 \}\), following circular, triangular, rectangular, or uniformly random layouts. These configurations define the sampling region \(\mathcal{R} \subseteq \mathcal{A}\), where noise floor measurements are collected. The jammer's position \( \mathbf{x}_j \) is also randomly assigned within \(\mathcal{A}\), independently of node placement, resulting in two distinct localization scenarios: when \( \mathbf{x}_j \in \mathcal{R} \), proximity to the source yields more precise sampling of interference, whereas when \( \mathbf{x}_j \in \mathcal{A} \setminus \mathcal{R} \), localization relies on extrapolation from peripheral observations.

Table~\ref{tab:rmse_fspl_ldpl_dataset_exp} presents the jammer localization performance across various methods in the static experiment as measured by \ac{RMSE}. \ac{MAE} results are provided in Appendix~\ref{app:further_results} Table~\ref{tab:mae_fspl_ldpl_dataset_exp}. The classical localization methods exhibit significantly higher errors across all sampling scenarios. Among them, \ac{WCL} performs best with an overall \ac{RMSE} of 143.1m, while \ac{LSQ} yields the worst performance. The suboptimal performance of path-loss-based approaches (\ac{PL}, \ac{MLE}, \ac{MLAT}, \ac{LSQ}) is likely attributed to their dependency on estimating the path loss exponent (\(\gamma\)) and jammer transmit power (\(P_t^{\text{jam}}\)) in order to estimate the jammer position~\cite{nardin2023apbm}. The \ac{GNN}-based approaches consistently outperform these classical methods. \ac{CAGE} delivers the highest overall performance, achieving an \ac{RMSE} of 66.6m, followed by \ac{GAT} with an \ac{RMSE} of 74.4m, improving localization accuracy as compared to \ac{WCL} by 114.9\%.

A significant performance gap is observed based on whether the jammer is located inside or outside the sampled region $\mathcal{R}$. Localization outside $\mathcal{R}$ is inherently more challenging due to extrapolation beyond measured signals. Classical methods like \ac{WCL} perform well within densely and radially symmetric sampled regions~\cite{wang2011weighted}, resulting in an \ac{RMSE} of 53.3m (inside). However, the performance sharply declines when extrapolating, with \ac{RMSE} increasing to 232.8m (outside). In contrast, \ac{GAT} demonstrates better robustness, with \ac{RMSE} rising moderately from 40.0m (inside) to 108.8m (outside). \ac{CAGE} achieves the lowest errors overall, with \ac{RMSE} of 35.7m (inside) and 97.5m (outside). Notably, the circular topology significantly enhances performance when the jammer is outside $\mathcal{R}$ by maximizing angular diversity and preserving directional information, as supported by the principles guiding the use of circular antenna arrays in direction-finding applications~\cite{hafner2019calibration}. These findings highlight the critical role of spatial coverage and the superior robustness of \ac{GNN}-based methods for long-range jammer localization.

\begin{table*}[t!]
    \centering
    \scriptsize
    \setlength{\tabcolsep}{0pt}
    \renewcommand{\arraystretch}{1.0}
    \caption{\ac{RMSE} in jammer localization along dynamic trajectories, reported across distance intervals to the jammer and averaged over three trials. The final column reports the mean RMSE across the full trajectory. MAE results are provided in Appendix~\ref{app:further_results} Table~\ref{tab:dynamic_path_exp_mae}.}
    \label{tab:dynamic_path_exp_rmse}
    \begin{tabularx}{\textwidth}{@{} p{0.4cm} p{1cm} *{5}{C} >{\hsize=1.25cm}C @{}}
    \toprule
     & \multirow{2}{*}{\textbf{Method}} & \multicolumn{5}{c}{\textbf{Distance to the Jammer (m)}} & \multirow{2}{*}{\textbf{Mean}} \\
    \cmidrule(lr){3-7}
         &        & $d > 500$ & $d \in [500,200]$ & $d \in [200,100]$ & $d \in [100,50]$ & $d \in [50,0]$ &   \\  
    \midrule
    \multirow{10}{*}{\rotatebox[origin=c]{90}{RMSE}} 
    & \ac{WCL} & 372.4 & 253.1 & 111.8 & 45.4 & 11.6 & 66.6 \\
    & \ac{PL} & 379.8 & 312.6 & 190.0 & 104.2 & 41.4 & 100.5 \\
    & \ac{MLE} & 510.5 & 463.9 & 148.4 & 92.9 & 6486.9 & 5493.5 \\
    & \ac{MLAT} & 275.2 & 234.3 & 183.9 & 147.0 & 99.1 & 124.7 \\
    \cmidrule(lr){2-8}
    & MLP & 182.1\sd{9.7} & 114.8\sd{7.8} & 49.1\sd{3.2} & 28.3\sd{1.2} & 19.0\sd{1.0} & 35.0\sd{0.7} \\
    & GCN & 161.5\sd{4.3} & 91.1\sd{1.2} & 34.2\sd{0.1} & 19.1\sd{0.4} & 10.8\sd{0.2} & 25.6\sd{0.3} \\
    & PNA & 214.1\sd{7.3} & 129.5\sd{5.2} & 56.9\sd{2.4} & 34.9\sd{1.7} & 24.9\sd{1.1} & 41.5\sd{1.2} \\
    & GAT & 131.0\sd{13.6} & 70.0\sd{0.2} & 33.2\sd{1.5} & 17.8\sd{0.6} & 9.6\sd{0.5} & 21.3\sd{0.6} \\
    & CAGE & \textbf{104.0\sd{5.9}} & \textbf{53.4\sd{0.2}} & \textbf{23.5\sd{0.7}} & \textbf{14.0\sd{0.4}} & \textbf{5.7\sd{0.2}} & \textbf{15.7\sd{0.1}} \\ 
    \bottomrule
    \end{tabularx}
\end{table*}

\subsection{Dynamic Evaluation for Jamming Localization}\label{sec:dynamic}
We evaluate localization performance in a dynamic 3D environment where a mobile device travels through space while tracking a jamming source. The trajectory follows an encirclement pattern, where the device converges toward the jammer while collecting measurements from varying distances and angles. 
This experiment enables the analysis of localization performance under varying observational constraints, evaluating how well methods estimate the jammer’s position with limited initial data and how accuracy evolves as additional spatially diverse measurements are incorporated. During training, we employ random cropping of trajectory segments to expose the model to varied subsequences of the jamming encounter, enhancing robustness to partial observations and improving generalization across different trajectory lengths.

\begin{figure}[!t]
    \centering
    \scriptsize
    \setlength{\tabcolsep}{2pt}
    \renewcommand{\arraystretch}{1.0}
    \begin{tabular}{cc}
        \includegraphics[width=0.475\textwidth]{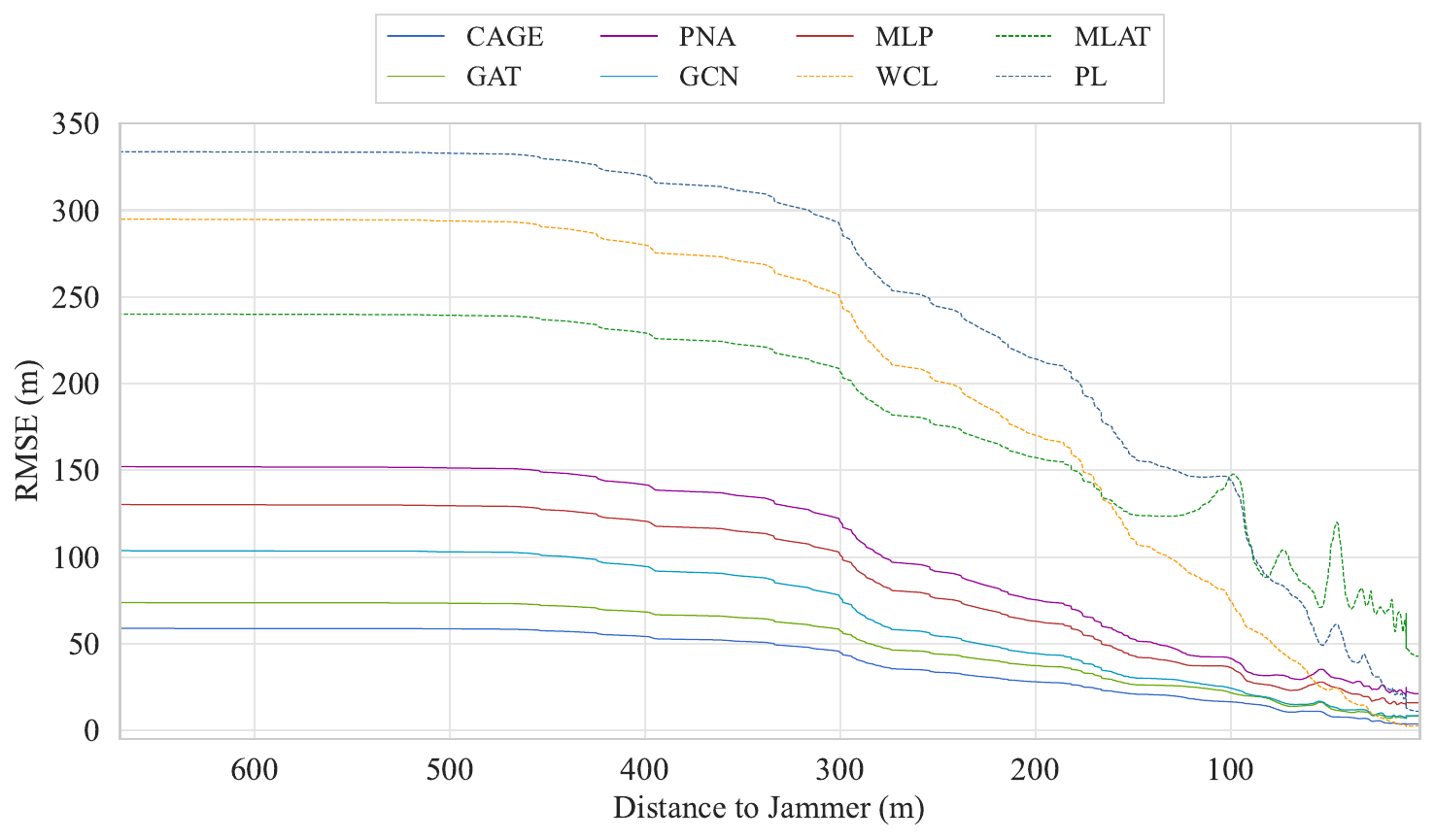} &
        \includegraphics[width=0.475\textwidth]{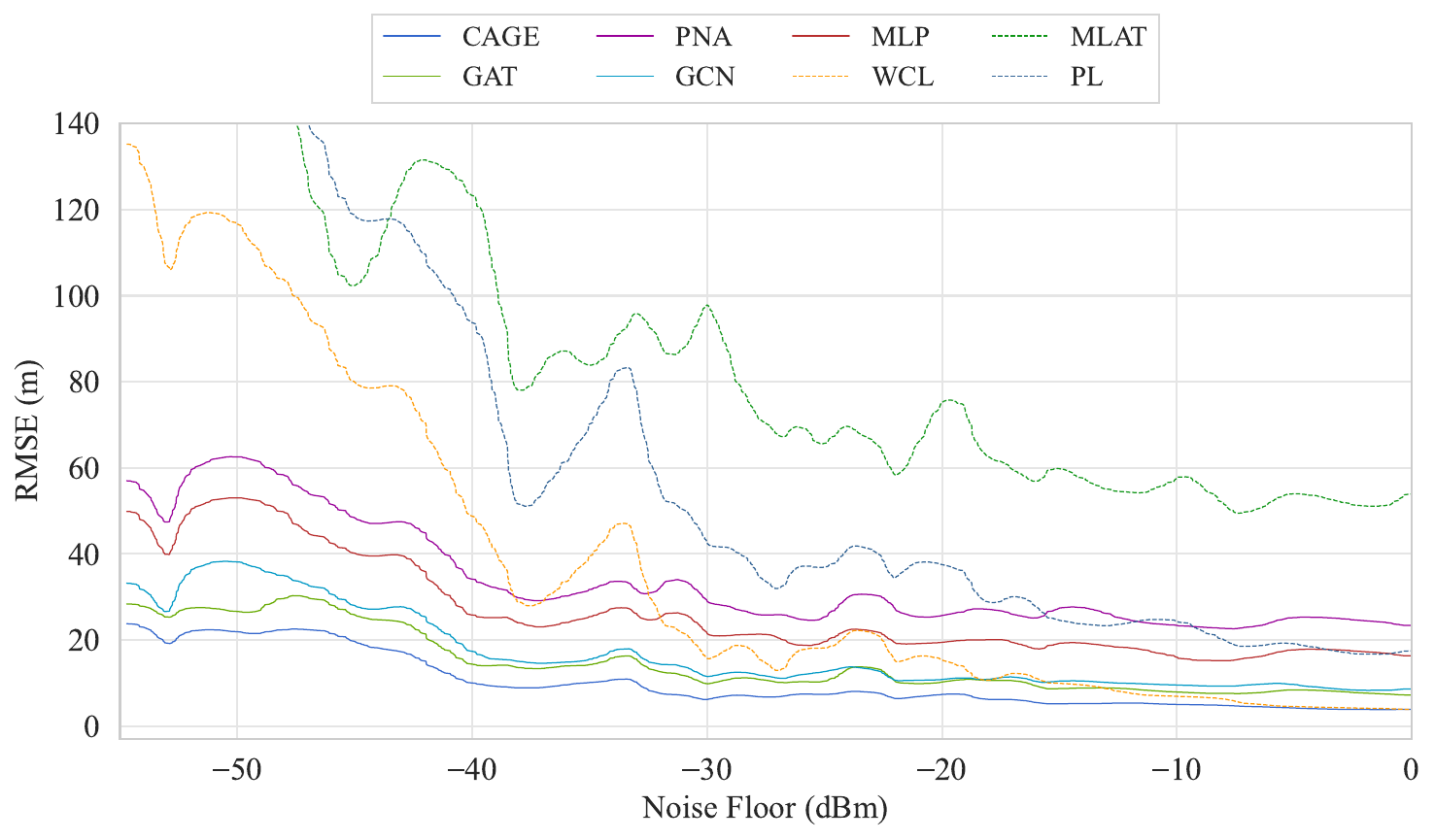} \\
        (a) Error vs. Min Distance to Jammer & (b) Error vs. Max Noise Floor
    \end{tabular}
    \caption{Localization performance as a function of (a) minimum distance to the jammer and (b) maximum noise floor.} 
    \label{fig:performance_trends}
\end{figure}

\begin{figure*}[!t]
    \centering
    \scriptsize
    \setlength{\tabcolsep}{1pt}
    \renewcommand{\arraystretch}{1.0}
    \begin{tabular}{c c c c c c}
        \includegraphics[width=0.16\textwidth]{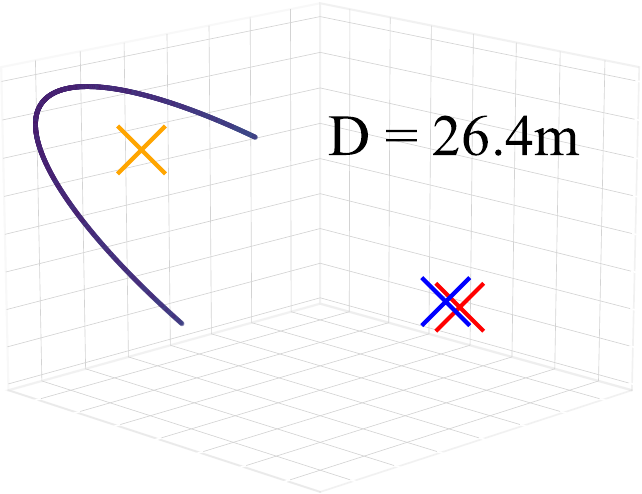} & 
        \includegraphics[width=0.16\textwidth]{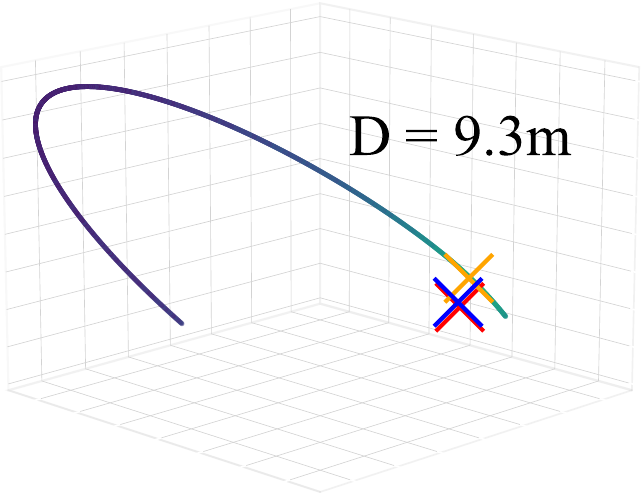} & 
        \includegraphics[width=0.16\textwidth]{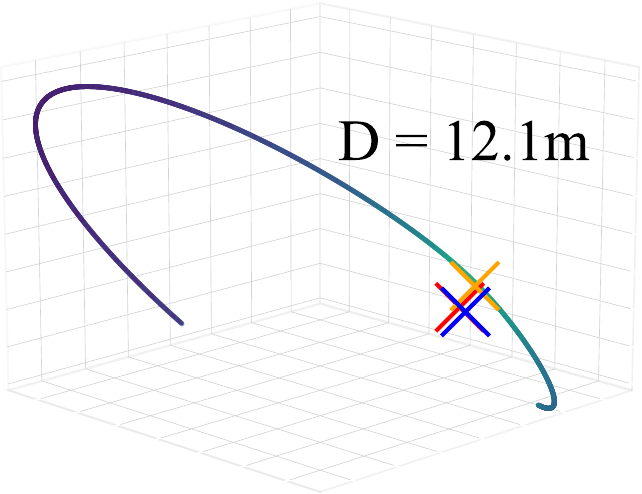} & 
        \includegraphics[width=0.16\textwidth]{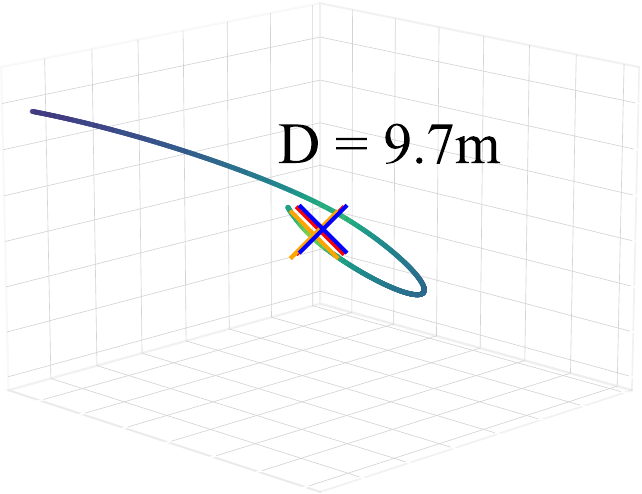} & 
        \includegraphics[width=0.16\textwidth]{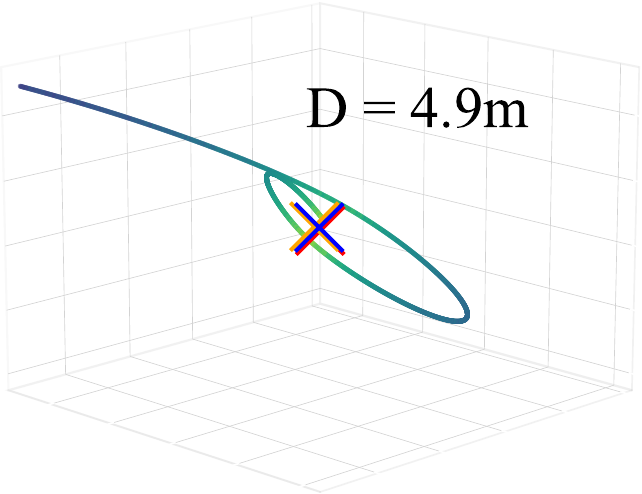} & 
        \includegraphics[width=0.16\textwidth]{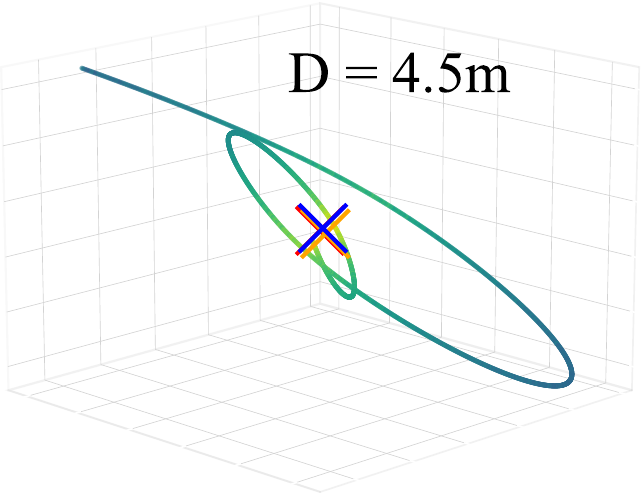} \\
        (a) $t=0$ & (b) $t=0.2$ & (c) $t=0.4$ & (d) $t=0.6$ & (e) $t=0.8$ & (f) $t=1.0$
    \end{tabular}
    \caption{Localization accuracy along a single trajectory. The red cross-mark is the jammer position, orange is WCL, and blue is the CAGE prediction.}
    \label{fig:trajectory_instance}
\end{figure*}

Table~\ref{tab:dynamic_path_exp_mae} presents the jammer localization performance for various methods in the dynamic experiment for \ac{MAE}. \ac{RMSE} results are provided in Appendix~\ref{app:further_results} Table~\ref{tab:dynamic_path_exp_rmse}. The results are categorized by distance intervals, where each column represents a different proximity range between the jammer and the nearest sampled measurement. The final column reports the overall mean across trajectories. Notably, WCL achieves an \ac{RMSE} of 11.6m in the closest range (\(d \in [50,0]\)). However, at larger distances (\(d > 500\)), WCL degrades significantly with an RMSE of 372.4m.  
\ac{CAGE} leverages confidence-weighted estimation to adaptively integrate WCL’s structured prior with the GNN-based predictions, resulting in the most consistent and robust performance. With an RMSE of 5.7m in \( d \in [50,0] \), CAGE significantly outperforms both WCL and GAT. Interestingly, while CAGE integrates WCL as a prior, its benefits extend beyond dense, symmetric scenarios. The performance gap is particularly evident in sparse sampling conditions, where CAGE maintains significantly lower error than WCL and GAT, highlighting its ability to generalize beyond the expected advantages of structured priors.

Figure~\ref{fig:performance_trends} shows localization performance trends over varying distances to the jammer and noise floor conditions. In Figure~\ref{fig:performance_trends}(a), WCL performs well at short distances, surpassing GAT only when very close to the jammer ($\sim$20\,m). While GAT demonstrates greater robustness along most of the trajectory, its performance is consistently surpassed by CAGE. Figure~\ref{fig:trajectory_instance} visually compares localization predictions along a trajectory, demonstrating progressive refinement after initial attack detection.

\subsection{Ablation Study}\label{sec:components}

\subsubsection{CAGE Components}
We perform an ablation study to assess the contribution of individual components in the CAGE architecture, as detailed in Table~\ref{tab:components}. The columns $\mathcal{L}_{\text{GNN}}$ and $\mathcal{L}_{\text{Adapt}}$ indicate the use of graph-based and adaptive losses, respectively, as defined in Section~\ref{sec:learning_framework}. ``SN'' and ``SN Edges'' denote the presence of a supernode and its edge connections. ``Con.'' represents the confidence weight ($\alpha$) layer (linear or 3-layer MLP with ReLU activations and final sigmoid), while ``Con. In'' and ``Reg. In'' indicate the input data for the confidence weight layer and regressor, respectively. ``Con. Out'' indicates whether a single shared confidence weight is used for all predicted coordinates or separate weights are produced for each. Finally, $\hat{\mathbf{x}}_{\text{final}}$ and $\hat{\mathbf{x}}_{\text{GNN}}$ represent the estimated position using the marked loss functions in the table and the GNN-only prediction, respectively. The symbols $\fullcircle$, $\halfcircle$, and $\emptycircle$ denote pooled graph representations with the supernode, without the supernode, and exclusively using the supernode. 

Results show that replacing bidirectional edges with directed edges improves accuracy, using an MLP instead of a linear layer reduces RMSE, and multiple confidence outputs outperform a single output. Training with \( L_{\text{CAGE}} \) enables the GNN regressor to function properly, achieving an RMSE of 16.2 compared to 153.0m without it, with adaptive estimation reaching 15.7.

\subsubsection{Confidence Weighting in CAGE}\label{app:confidence}
We analyze how the \( \mathcal{L}_{\text{CAGE}} \) loss shapes confidence weight assignment in CAGE by comparing two training setups: one using only \( \mathcal{L}_{\text{Adapt}} \) and another incorporating the full \( \mathcal{L}_{\text{CAGE}} \) loss. Confidence weights are plotted against the jammer’s distance to assess how the choice of loss function influences the model's reliance on the WCL prior versus GNN-based predictions. This experiment uses the dynamic trajectory data described in Section~\ref{sec:dynamic}, where measurements are collected while encircling the jammer from varying distances and angles. This setup yields conditions where the GNN can best leverage wide spatial context at long ranges, while WCL becomes increasingly reliable at close distances due to the dense, radially symmetric sampling created by the encircling motion.

\begin{table*}[!t]
    \centering
    \scriptsize
    \setlength{\tabcolsep}{0pt}
    \renewcommand{\arraystretch}{1.0}
    \caption{Ablation study on CAGE components.}
    \label{tab:components}
    \begin{tabularx}{\textwidth}{@{} >{\hsize=0.75cm}C>{\hsize=0.75cm}C *{5}{C} C C C   @{}}
    \toprule
    \multicolumn{2}{c}{\textbf{Loss}} & \multicolumn{5}{c}{\textbf{Configurations}} & & \multirow{2}{*}{$\hat{\mathbf{x}}_{\text{final}}$} & \multirow{2}{*}{$\hat{\mathbf{x}}_{\text{GNN}}$} \\
    \cmidrule(lr){1-2} \cmidrule(lr){3-8}  
    $\mathcal{L}_{\text{GNN}}$ & $\mathcal{L}_{\text{Adapt}}$ & \textbf{SN} & \textbf{SN Edges} & \textbf{Con.} & \textbf{Con. In} & \textbf{Reg. In} & \textbf{Con. Out} & \\
        \midrule
        \cmark & \xmark & \xmark & -- & -- & -- & \halfcircle & -- & -- & 21.3\sd{0.6} \\ 
        \xmark & \cmark & \cmark & Undirected & Linear & \emptycircle & \halfcircle & Single & 22.2\sd{0.2} & 253.8\sd{40.1} \\
        \xmark & \cmark & \cmark & Undirected & Linear & \fullcircle & \fullcircle & Single & 21.3\sd{0.7} & 247.4\sd{30.8}\\
        \xmark & \cmark & \cmark & No edges & Linear & \fullcircle & \fullcircle & Single & 20.6\sd{0.1} & 155.8\sd{32.2}\\
        \xmark & \cmark & \cmark & No edges & Linear & \emptycircle & \halfcircle & Single & 19.7\sd{0.3} & 119.1\sd{9.2} \\ 
        \xmark & \cmark & \cmark & Directed & Linear & \emptycircle & \halfcircle & Single & 19.4\sd{0.3} & 102.9\sd{6.9} \\ 
        \xmark & \cmark & \cmark & Directed & MLP & \emptycircle & \halfcircle & Single &  18.2\sd{1.1} & 117.1\sd{4.7}\\ 
        \xmark & \cmark & \cmark & Directed & MLP & \emptycircle & \halfcircle & Multiple & 17.7\sd{0.4} & 153.0\sd{17.1}\\ 
        \cmark & \cmark & \cmark & Directed & MLP & \emptycircle & \halfcircle & Single & 16.6\sd{0.8} & 16.8\sd{0.8} \\ 
        \cmark & \cmark & \cmark & Directed & MLP & \emptycircle & \halfcircle & Multiple & 15.7\sd{0.1} & 16.2\sd{0.0} \\ 
        \bottomrule
    \end{tabularx} 
\end{table*}

\begin{figure}[t]
    \centering
    \scriptsize
    \setlength{\tabcolsep}{2pt}
    \renewcommand{\arraystretch}{1.0}
    \begin{tabular}{cc}
        \includegraphics[width=0.475\textwidth]{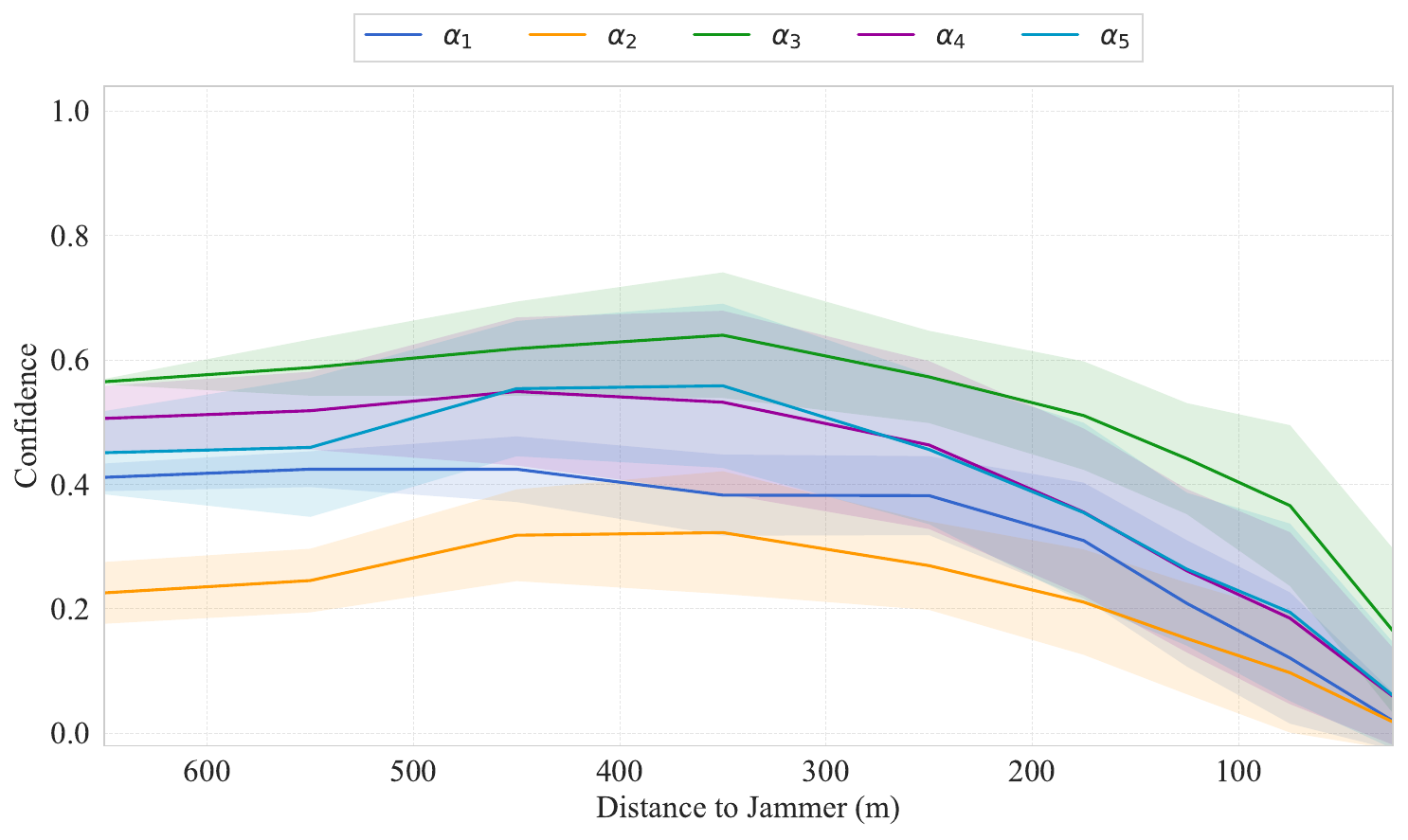} & 
        \includegraphics[width=0.475\textwidth]{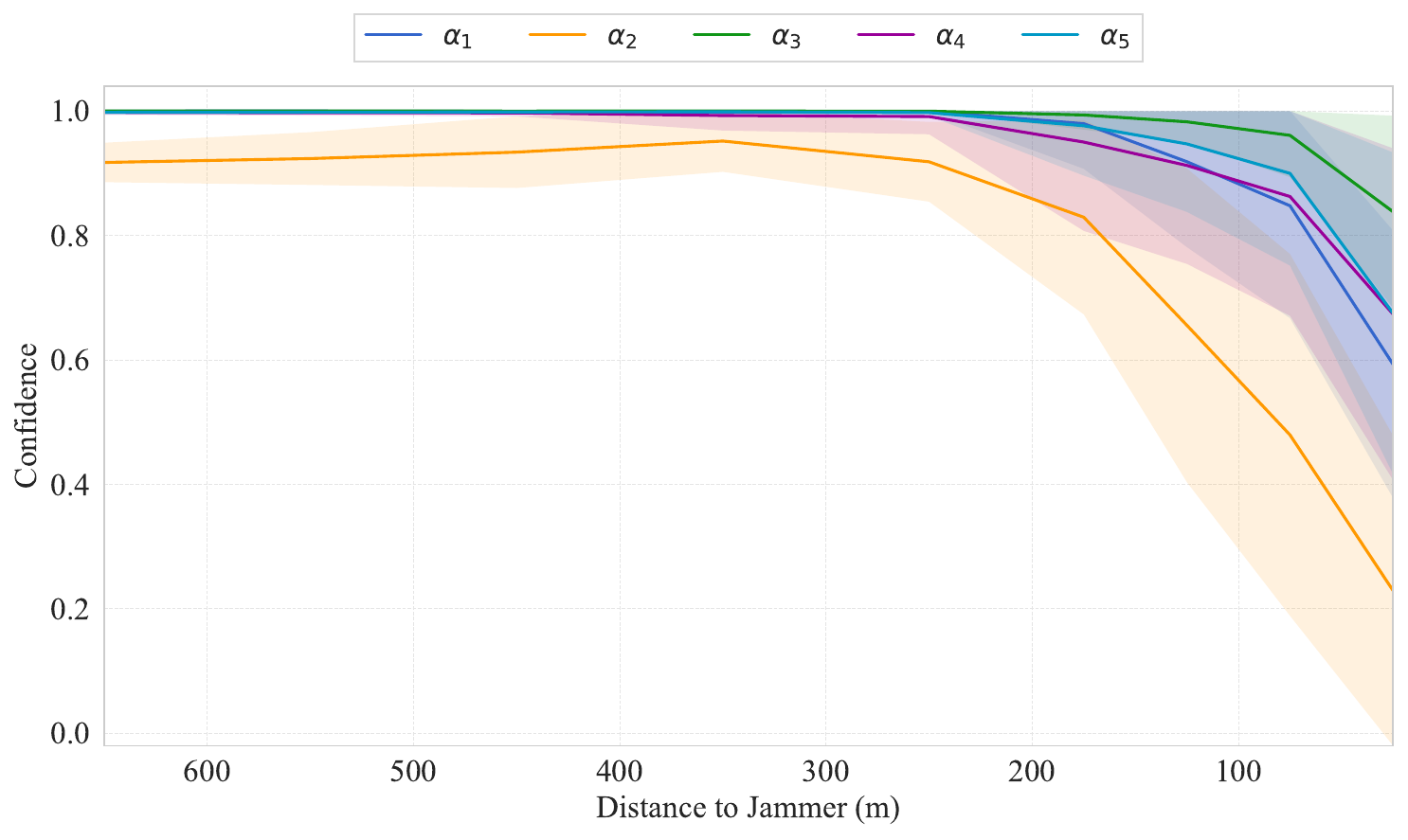} \\ 
        (a) Training with $\mathcal{L}_{\text{Adapt}}$ & (b) Training with $\mathcal{L}_{\text{CAGE}}$
    \end{tabular}
    \caption{Effect of training loss on confidence weighting in CAGE. (a) Training with $\mathcal{L}_{\text{Adapt}}$ causes over-reliance on WCL due to low GNN confidence. (b) Training with $\mathcal{L}_{\text{CAGE}}$ increases initial GNN confidence, sharpening near the jammer.}
    \label{fig:confidence_analysis}
\end{figure}

When trained solely with $\mathcal{L}_{\text{Adapt}}$ (Figure~\ref{fig:confidence_analysis}a), the model assigns relatively low confidence weights across all distances. This suggests a persistent reliance on WCL, even in regions where GNN-based predictions should dominate. Since $\mathcal{L}_{\text{Adapt}}$ optimizes the blended prediction without enforcing a direct learning signal for the GNN, the model tends to treat it as a correction mechanism rather than as an independent estimator. In contrast, training with $\mathcal{L}_{\text{CAGE}}$ (Figure~\ref{fig:confidence_analysis}b) yields high initial confidence in the GNN prediction ($\alpha \approx 1$), especially at large distances. As the distance to the jammer decreases, confidence in WCL increases, indicating that the model learns when the WCL prior becomes more reliable, particularly in the densely sampled, radially symmetric regions near the jammer. This improvement stems from the explicit GNN supervision provided by $\mathcal{L}_{\text{GNN}}$, resulting in a more structured confidence-weighting mechanism and improved localization robustness.

Figure~\ref{fig:confidence_analysis}b also reveals that among all confidence weights, $\alpha_2$, which corresponds to $\sin(\theta)$ in the final estimation, remains slightly lower than the others, even at large distances. This suggests that while the model generally prioritizes GNN-based predictions, it continues to rely on the WCL azimuth as a stable directional reference. Although we used $\lambda = 0$ in this work, tuning $\lambda$ in Equation~\eqref{eq:loss_cage} may further improve performance by better balancing the model’s reliance on different components.

\begin{table*}[t]
    \centering
    \scriptsize
    \setlength{\tabcolsep}{3pt}
    \renewcommand{\arraystretch}{1.0}
    \caption{Ablations on augmentations, global pooling, and neighborhood size.}
    \label{tab:combined_ablation}
    \begin{tabular}{cc}
        (a) Varying number of \textit{k} nearest neighbors. & (b) Global pooling strategies. \\ 
        \begin{tabularx}{0.48\textwidth}{XCCCC}
            \toprule
            $\mathbf{k}$ & \textbf{3} & \textbf{5} & \textbf{7} & \textbf{11} \\
            \midrule
            RMSE & \textbf{54.7\sd{0.1}} & 57.3\sd{0.0} & 60.2\sd{0.2} & 64.9\sd{0.2} \\
            \bottomrule
        \end{tabularx} 
        &
        \begin{tabularx}{0.48\textwidth}{XCCCC}  
            \toprule
            \textbf{Pooling} & \textbf{Sum} & \textbf{Mean} & \textbf{Max} & \textbf{Att} \\
            \midrule
            RMSE & 72.2\sd{0.3} & 84.1\sd{0.2} & \textbf{52.9\sd{0.1}} & 61.7\sd{0.3} \\
            \bottomrule
        \end{tabularx}  
    \end{tabular}
    
    \vspace{1em} 
    (c) Graph- and Feature-level augmentations. \\
    \begin{tabularx}{0.9\textwidth}{XCCCCC} 
        \toprule
        \textbf{Aug} & \textbf{None} & \textbf{Rotation} & \textbf{Crop} & \textbf{Drop Node} & \textbf{Feat. Noise} \\
        \midrule
        RMSE & 54.7\sd{0.1} & 58.9\sd{0.1} & 54.5\sd{0.1} & \textbf{52.9\sd{0.1}} & 54.6\sd{0.1} \\
        \bottomrule
    \end{tabularx} 

    \vspace{1em} 
    (d) Performance of combined augmentation strategies. \\
    \begin{tabularx}{0.9\linewidth}{XCCC}
        \toprule
        \textbf{Aug} & \textbf{Crop+DN} & \textbf{DN+Feat Noise} & \textbf{Crop+Feat Noise} \\
        \midrule
        RMSE & 55.5\sd{0.1} & \textbf{53.0\sd{0.1}} & 55.3\sd{0.2} \\
        \bottomrule
    \end{tabularx}
\end{table*}

\subsubsection{Graph Construction, Pooling, and Augmentations}  
We investigate how different graph construction strategies, pooling methods, and augmentation techniques influence model performance, as summarized in Table~\ref{tab:combined_ablation}. A lower neighborhood size (\( k=3 \)) yields the best RMSE, with larger \( k \) leading to oversmoothing~\cite{li2018oversmoothing}. Among global pooling strategies, max pooling performs best. For augmentations, DropNode~\cite{do2021dropnode} at a 0.2 drop rate offers the greatest improvement, and the combination of DropNode and feature noise achieves the best performance overall. See Appendix~\ref{app:ablations} for further results on feature engineering and augmentations.

\subsubsection{Downsampling Strategies for High-Resolution Signal Graphs}\label{app:downsampling}
High-frequency sampling in \ac{SDR} systems produces dense graphs with thousands of nodes (approx.\ 6000 per instance in our dynamic dataset; see Appendix~\ref{app:modeling}, Table~\ref{tab:dataset_statistics}). To reduce computational cost while preserving key signal characteristics, we apply downsampling prior to graph construction. We compare two methods: (1) \textit{window averaging}, which operates on the raw sequence of samples by dividing it into \(|V|\) segments based on sample count and averaging the position and noise values within each; and  (2) \textit{spatial binning with noise filtering}, which groups samples into fixed 1m\(^3\) spatial bins based on their positions, averages the position and noise values within each bin, and retains only the \(|V|\) bins with the highest average noise, as high-noise regions are more informative~\cite{nardin2023apbm}.

\begin{table*}[t]
    \centering
    \scriptsize
    \setlength{\tabcolsep}{0pt}
    \renewcommand{\arraystretch}{1.0}
    \caption{Ablation study of downsampling techniques on high-resolution dynamic path data under LDPL model. The retained number of nodes is denoted by $|V|$.}\label{tab:downsampling_ablation}
    \begin{tabularx}{\textwidth}{@{} X *{4}{C} *{4}{C} @{}}
    \toprule
    \multirow{2}{*}{$|V|$} & \multicolumn{4}{c}{Window averaging} & \multicolumn{4}{c}{Spatial binning with noise filtering} \\
    \cmidrule(lr){2-5} \cmidrule(lr){6-9}
        & $t_{0.0-0.2}$ & $t_{0.4-0.6}$ & $t_{0.8-1.0}$ & Mean & $t_{0.0-0.2}$ & $t_{0.4-0.6}$ & $t_{0.8-1.0}$ & Mean \\
    \midrule
    200 & 74.5 & 22.1 & 13.2 & 38.7 & 78.4 & 21.6 & 10.5 & 40.5 \\
    600 & 63.2 & 21.8 & 14.7 & 34.3 & 63.5 & 16.3 & 9.8 & 31.7 \\
    800 & 57.1 & 23.9 & 15.7 & 32.2 & 59.3 & 15.3 & 10.4 & 29.7 \\
    1000 & 54.6 & 23.4 & 16.5 & \textbf{31.4} & 54.5 & 13.7 & 9.8 & \textbf{27.4} \\
    \bottomrule
    \end{tabularx}
\end{table*}

Table~\ref{tab:downsampling_ablation} presents results from the downsampling ablation experiment. Each segment \( t_{\ell,u} \) denotes a subinterval of the normalized trajectory, with \( \ell \) and \( u \) indicating the lower and upper bounds of the time fraction (e.g., \( t_{0.8,1.0} \) corresponds to the final 20\% of the path). The results indicate that spatial binning with noise filtering consistently outperforms window averaging, particularly in later segments. Following these results, we adopt the spatial boning with noise filtering with $|V| = 1000$ for all our experiments.

\section{Conclusion}
This work presents CAGE, the first graph-based framework for jamming source localization that reformulates the problem as an inductive graph regression task. Our approach integrates attention-based GNNs with a confidence-weighted fusion mechanism that adaptively balances learned predictions with structured spatial priors. Experiments across diverse static and dynamic scenarios demonstrate that CAGE consistently outperforms both classical and learning-based baselines. Through detailed ablations, we show that our design choices significantly enhance model robustness and performance. Future work includes incorporating temporal GNNs, analyzing the impact of the confidence weighting parameter \(\lambda\), and evaluating cross-domain generalization beyond simulated environments.

\bibliographystyle{splncs04}
\bibliography{mybibliography} 

\clearpage\newpage
\appendix
\section{Ablation Studies}\label{app:ablations}

We conduct a series of ablation studies to evaluate the impact of different design choices and components on localization performance.

\subsection{Node Features, Graph Construction, and Augmentations}  
Table~\ref{tab:node_feats_combined} evaluates the impact of various node features in static and dynamic settings using the GCN model on a validation set. Incorporating Cartesian coordinates and weighted centroid distances improves RMSE, while azimuth-based features do not provide benefits. Local noise statistics, particularly mean noise deviation, significantly enhance accuracy. Furthermore, we incorporate unique IDs~\cite{sato2021random}, with which we observe a degradation in performance, possibly due to a loss in the generalization ability of the model~\cite{zhang2021nested}. In dynamic scenarios, direction vectors and temporal noise variations further reduce RMSE, highlighting the importance of motion-aware attributes.  
\begin{table*}[h]
    \centering
    \scriptsize
    \setlength{\tabcolsep}{3pt}
    \renewcommand{\arraystretch}{1.0}
    \caption{Node feature ablation study.}
    \label{tab:node_feats_combined}
    \begin{tabular}{cc} 
        (a) Static. & (b) Dynamic. \\
        \begin{tabularx}{0.48\textwidth}{@{} >{\hsize=3.5cm}X C @{}} 
            \toprule
            \textbf{Node Feature} & \textbf{RMSE (m)} \\ 
            \midrule
            Baseline & 67.4\sd{0.1} \\
            + $\mathbf{x}_i^{\text{cart}}$ & 66.0\sd{0.1} \textcolor{mygreen}{$\downarrow$} \textcolor{mygreen}{\scriptsize 2.1\%} \\
            + $\ d_i^{\text{wcent}}$ & 66.0\sd{0.1}  - {0.0 \scriptsize \%} \\ 
            + $\mathbf{x}_i^{\text{wcent}}$ & 63.6\sd{0.2} \textcolor{mygreen}{$\downarrow$} \textcolor{mygreen}{\scriptsize 3.6\%}\\ 
            + Azimuth to centroid & 66.1\sd{0.2} \textcolor{myred}{$\uparrow$} \textcolor{myred}{\scriptsize 0.2\%} \\ 
            + $\ d_i^{\text{wcent}}$ & 63.3\sd{0.2} \textcolor{mygreen}{$\downarrow$} \textcolor{mygreen}{\scriptsize 0.5\%}\\ 
            + Azimuth to WC & 65.5\sd{0.5}  \textcolor{myred}{$\uparrow$} \textcolor{myred}{\scriptsize 3.5\%}\\ 
            + $\text{median}(\eta_{\mathcal{N}_k(i)})$ & 61.3\sd{0.2}  \textcolor{mygreen}{$\downarrow$} \textcolor{mygreen}{\scriptsize 3.2\%} \\
            + $\max(\eta_{\mathcal{N}_k(i)})$ & 59.4\sd{0.2}  \textcolor{mygreen}{$\downarrow$} \textcolor{mygreen}{\scriptsize 3.2\%} \\
            + $\Delta\eta_i$ & 57.3\sd{0.0}  \textcolor{mygreen}{$\downarrow$} \textcolor{mygreen}{\scriptsize 3.5\%} \\
            + Random feature & 57.5\sd{0.1}  \textcolor{myred}{$\uparrow$} \textcolor{myred}{\scriptsize 0.3\%} \\
            \bottomrule
        \end{tabularx} 
        &
        \begin{tabularx}{0.48\textwidth}{@{} >{\hsize=3.5cm}X C @{}} 
            \toprule
            \textbf{Node Feature} & \textbf{RMSE (m)}  \\ 
            \midrule
            Baseline & 27.4\sd{0.3} \\
            + Moving avg. signal & 27.5\sd{0.3} \textcolor{myred}{$\uparrow$} \textcolor{myred}{\scriptsize \,0.4\%} \\
            + Path distance & 32.7\sd{0.4} \textcolor{myred}{$\uparrow$} \textcolor{myred}{\scriptsize 19.3\%} \\
            + $\mathbf{d}_i$ & 26.2\sd{0.2} \textcolor{mygreen}{$\downarrow$} \textcolor{mygreen}{\scriptsize \,4.4\%} \\
            + $\Delta \eta_{i}^\text{temp}$ & 25.6\sd{0.3} \textcolor{mygreen}{$\downarrow$} \textcolor{mygreen}{\scriptsize \,6.6\%} \\
            \bottomrule
        \end{tabularx}  
    \end{tabular}
\end{table*}

\subsection{Data Augmentations}
In this section, we present additional experiments related to data augmentations, including feature corruption, rotation augmentation, DropNode~\cite{do2021dropnode}, and random cropping. 
Feature corruption~\cite{ding2022gda} applies Gaussian noise to a randomly selected subset of node features. Up to half of the total features in each vector are chosen, and a mask determines which node values receive noise with probability \( p \). Rotation augmentation rotates the graph by a random angle between 0 and 360 degrees to encourage orientation invariance. DropNode removes nodes with a drop rate of \( p \). The scaling factor is deliberately excluded to maintain the integrity of physical relationships. Random cropping selects a portion of the graph by identifying the three nodes with the highest noise levels and extending to a randomly chosen endpoint. Although we recognize the preexisting GraphCrop augmentation\cite{wang2020graphcrop}, this method is not relevant for our application which is characterized by uniform node degree and localized areas of importance in the graph. Table~\ref{tab:combined_ablation} presents the results of these experiments, showing that the best performance is achieved with DropNode at \( p = 0.2 \). Consequently, we adopt this setting exclusively during training, as discussed in Section~\ref{sec:results}.
\begin{table}[h]
    \centering
    \scriptsize
        \caption{Performance comparison of different augmentation techniques.}\label{tab:aug_p_ablation}
    \begin{tabularx}{\columnwidth}{XCC} 
        \toprule
        Augmentation & \( p \) & RMSE \\
        \midrule
        Feature Noise & 1.0 & 55.0\sd{0.1}\\
        Feature Noise & 0.5 & 54.6\sd{0.1}\\
        Feature Noise & 0.1 & 54.7\sd{0.1}\\
        \midrule
        DropNode & 0.5 & 54.7\sd{0.2}\\
        DropNode & 0.2 & \textbf{52.9\sd{0.1}}\\ 
        DropNode & 0.1 & 53.3\sd{0.1}\\ 
        \bottomrule
    \end{tabularx}
\end{table}

We also evaluate combined augmentation strategies where Crop+DN refers to cropping followed by DropNode, DN+Feat Noise represents DropNode followed by feature corruption, and Crop+Feat Noise denotes cropping followed by feature corruption. Table~\ref{tab:combined_ablation}d illustrates that combining these augmentations does not significantly improve performance. The best result is achieved with DN + Feat Noise, reinforcing our decision to use DropNode with \( p = 0.2 \) exclusively.

\begin{table*}[h]
    \centering
    \scriptsize
    \setlength{\tabcolsep}{0pt}
    \renewcommand{\arraystretch}{1.0}
    \caption{\ac{MAE} in jammer localization for static scenarios, averaged over three trials with different seeds. Results are split by sampling geometry.}
    \label{tab:mae_fspl_ldpl_dataset_exp} 
    \begin{tabularx}{\textwidth}{@{} p{0.25cm} p{1cm} *{5}{C} *{5}{C} p{0.75cm}@{}}
    \toprule
    & \multirow{2}{*}{\textbf{Method}} & \multicolumn{5}{c}{\textbf{Jammer within ($\mathbf{x}_j \in \mathcal{R}$)}}  
    & \multicolumn{5}{c}{\textbf{Jammer outside ($\mathbf{x}_j \in \mathcal{A} \setminus \mathcal{R} $)}} 
    & \multirow{2}{*}{\textbf{Mean}} \\
    \cmidrule(lr){3-7} \cmidrule(lr){8-12}
    & & \textbf{C}  & \textbf{T} & \textbf{R} & \textbf{RD} & \textbf{Mean} 
      & \textbf{C} & \textbf{T} & \textbf{R} & \textbf{RD} & \textbf{Mean} & \\
    \midrule
    \multirow{10}{*}{\rotatebox[origin=c]{90}{MAE}} 
    & \, \ac{WCL} & 37.7 & 40.2 & 27.2 & 35.7 & 35.2 & 138.6 & 169.0 & 170.1 & 164.2 & 160.5 & 97.9 \\
    & \, \ac{PL} & 122.2 & 83.3 & 90.6 & 82.0 & 94.5 & 289.4 & 247.3 & 290.6 & 267.4 & 273.7 & 184.1 \\ 
    & \, MLE & 70.5 & 65.5 & 63.4 & 67.4 & 66.7 & 208.4 & 197.9 & 203.5 & 230.3 & 210.0 & 138.4 \\ 
    & \, \ac{MLAT} & 121.6 & 89.6 & 75.3 & 64.8 & 87.8 & 282.2 & 259.9 & 275.2 & 235.8 & 263.3 & 175.6 \\ 
    & \, \ac{LSQ} & 237.9 & 138.5 & 92.0 & 94.1 & 140.6 & 410.5 & 329.7 & 351.1 & 310.9 & 350.6 & 245.6 \\ 
    \cmidrule(lr){2-13}
    & \, \ac{MLP} & 37.1 & 31.3 & 24.7 & 29.5 & 30.7 & 60.1 & 78.2 & 78.5 & 84.1 & 75.2 & 53.0 \\  
    & \, \ac{GCN} & 35.5 & 30.7 & 25.9 & 35.3 & 31.9 & 57.5 & 75.5 & 77.4 & 85.4 & 74.0 & 53.0 \\  
    & \, PNA & 34.3 & 27.7 & 21.9 & 26.7 & 27.7 & 56.5 & 72.7 & 74.1 & 79.4 & 70.7 & 49.2 \\ 
    & \, \ac{GAT} & 33.3 & 27.4 & 21.5 & 27.2 & 27.4 & 55.8 & 72.0 & 73.5 & 77.9 & 69.8 & 48.6 \\  
    & \, CAGE & \textbf{28.4} & \textbf{23.6} & \textbf{19.8} & \textbf{25.0} & \textbf{24.2} & \textbf{46.5} & \textbf{61.6} & \textbf{65.1} & \textbf{68.8} & \textbf{60.5} & \textbf{42.5} \\  
    \bottomrule
    \end{tabularx}
\end{table*}

\begin{table*}[h]
    \centering
    \scriptsize
    \setlength{\tabcolsep}{0pt}
    \renewcommand{\arraystretch}{1.0}
    \caption{\ac{MAE} in jammer localization along dynamic trajectories, reported across distance intervals to the jammer and averaged over three trials. The final column reports the mean MAE across the full trajectory.}
    \label{tab:dynamic_path_exp_mae}
    \begin{tabularx}{\textwidth}{@{} p{0.4cm} p{1cm} *{5}{C} >{\hsize=1.25cm}C @{}}
    \toprule
     & \multirow{2}{*}{\textbf{Method}} & \multicolumn{5}{c}{\textbf{Distance to the Jammer}} & \multirow{2}{*}{\textbf{Mean}} \\
    \cmidrule(lr){3-7}
         &        & $d > 500$ & $d \in [500,200]$ & $d \in [200,100]$ & $d \in [100,50]$ & $d \in [50,0]$ &   \\  
    \midrule
    \multirow{10}{*}{\rotatebox[origin=c]{90}{MAE}} 
    & \ac{WCL} & 316.8 & 220.6 & 95.2 & 34.6 & 7.7 & 28.2 \\
    & \ac{PL} &  323.8 & 275.3 & 153.5 & 74.0 & 28.9 & 56.3\\
    & \ac{MLE} & 245.1 & 130.0 & 54.2 & 36.5 & 1381.8 & 1006.7 \\
    & \ac{MLAT} & 237.8 & 192.5 & 122.2 & 85.8 & 57.7 & 73.1 \\
    \cmidrule(lr){2-8}
    & MLP & 140.5\sd{9.7} & 83.7\sd{4.9} & 35.5\sd{1.5} & 21.8\sd{0.9} & 14.8\sd{0.8} & 20.7\sd{0.4} \\
    & GCN & 123.0\sd{3.8} & 65.4\sd{1.0} & 25.2\sd{0.1} & 14.3\sd{0.2} & 8.1\sd{0.2} & 13.1\sd{0.1} \\
    & PNA & 162.1\sd{5.2} & 95.1\sd{2.6} & 40.9\sd{2.2} & 26.3\sd{1.1} & 19.5\sd{1.0} & 25.7\sd{1.0} \\
    & GAT & 88.7\sd{10.6} & 49.3\sd{1.4} & 23.1\sd{0.7} & 13.3\sd{0.4} & 7.3\sd{0.4} & 11.4\sd{0.5} \\
    & CAGE & \textbf{73.1\sd{0.7}} & \textbf{36.7\sd{0.9}} & \textbf{17.1\sd{0.6}} & \textbf{9.9\sd{0.1}} & \textbf{4.2\sd{0.1}} & \textbf{7.6\sd{0.1}} \\ 
    \bottomrule
    \end{tabularx}
\end{table*}

\section{Further Results and Analysis}\label{app:further_results}
\begin{figure*}[h]
    \centering
    \includegraphics[width=1.0\textwidth]{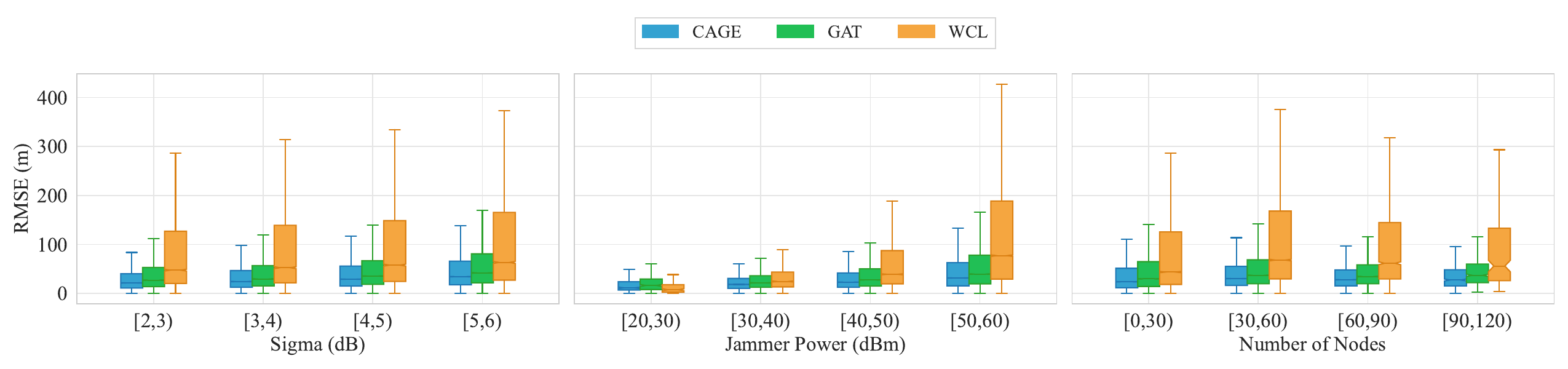}
    \caption{RMSE analysis of the CAGE, GAT and WCL models across all topologies under the LDPL environment with varying shadowing effects, jammer transmit powers, and number of nodes per graph.}
    \label{fig:combined_rmse_analysis}
\end{figure*}

Figure~\ref{fig:combined_rmse_analysis} illustrates the impact of varying parameters, such as jammer power, network density, and shadowing effects, on localization performance for \ac{CAGE}, \ac{GAT}, and \ac{WCL} on the static experiment data. Across all conditions, \ac{CAGE} consistently achieves the lowest \ac{RMSE}, highlighting its robustness. Stronger shadowing (\(\sigma\)) degrades performance across all methods due to increased signal uncertainty~\cite{tedeschi2022pwrmodjam}. An inverse relationship between jammer power and localization accuracy emerges, likely because lower power forces nodes closer to the jammer, improving localization (see Tables~\ref{tab:mae_fspl_ldpl_dataset_exp} and \ref{tab:rmse_fspl_ldpl_dataset_exp}). As shown in Appendix~\ref{app:modeling}, Table~\ref{tab:dataset_statistics}, the 2D graphs predominantly consists of instances with a small number of nodes. This imbalance limits the model's exposure to such graphs, leading to degraded performance on larger graphs. The lack of sufficient representation may contribute to feature explosions, a phenomenon consistent with prior findings~\cite{corso2020pna}.

\newpage
\section{Training Details and Hyperparameter Tuning}\label{app:hyperparams_modelarch}

Each model is trained for 300 epochs using the AdamW optimizer, with weights initialized via Xavier initialization~\cite{xavier}. The learning rate is linearly warmed up over the first $20\%$ of the total epochs and then follows a cosine annealing schedule. Dropout is set to 0.5~\cite{dropout} and is applied to graph representations before the final linear layers for all models, except for CAGE, where it is set to 0, as this configuration was found to yield the best results. Weight decay is fixed at $10^{-5}$ for all models.

We perform hyperparameter tuning using the \ac{TPE} sampler and Hyperband pruner for \ac{MLP}, \ac{GCN}, and \ac{GAT}. \ac{PNA} hyperparameters follow~\cite{chen2022structure}, utilizing mean, maximum, and standard deviation as node feature aggregators. The minimum aggregator was excluded as it degraded performance. The hyperparameters for each model are summarized in Table~\ref{tab:hyperparams_models}, where ``LR'' denotes learning rate, ``WD'' refers to weight decay, ``BS'' represents batch size, ``DO'' indicates dropout rate, ``NH'' corresponds to the number of attention heads (for applicable models), ``NL'' is the number of layers, and ``HC/OC'' represents the number of hidden and output channels, respectively.

\begin{table*}[!htbp]
    \centering
    \scriptsize
    \setlength{\tabcolsep}{0pt}
    \renewcommand{\arraystretch}{1.0}
    \caption{Model hyperparameter tuning results, detailing the optimized learning rates (LR), weight decay (WD), batch sizes (BS), dropout rates (DO), number of attention heads (NH), number of layers (NL), and hidden/output channels (HC/OC) for each model.}
    \label{tab:hyperparams_models}
    \begin{tabularx}{\textwidth}{XCCCCCCC}
    \toprule
    Model & LR & WD & BS & DO & NH & NL & HC/OC \\
    \midrule
    MLP & 0.0004 & 0.00001 & 16 & 0.5 & -- & 8 & 128 / 64 \\
    GCN & 0.0005 & 0.00001 & 32 & 0.5 & -- & 2 & 512 / 256 \\
    PNA & 0.001 & 0.00001 & 128 & 0.5 & -- & 6 & 64 / 64 \\
    GAT & 0.0007 & 0.00001 & 8 & 0.5 & 4 & 8 & 128 / 128 \\
    \bottomrule
    \end{tabularx} 
\end{table*}

\newpage
\section{Signal Propagation and Interference Modeling} \label{app:modeling}

This section complements the problem definition by describing how the \ac{RSSI}, noise floor, and jammer interference are modeled to generate instances for the jammer localization problem. The environment modeling incorporates signal propagation models and the effects of the jammer on the network.

\subsubsection{RSSI Modeling}
The RSSI (\(P_{\text{rx},i}\)) at a receiving device \( i \) from a transmitting device is calculated as:
\[
P_{\text{rx},i} = P_t + G_t + G_r - \text{PL}(d_i),
\]
where \(P_t\) is the transmit power of the transmitting device (in dBm), \(G_t\) and \(G_r\) are the antenna gains of the transmitting and receiving devices (in dBi), and \(\text{PL}(d_i)\) is the path loss (in dB) as defined in Equation~\eqref{eq:LDPL}, determined by the distance \(d_i\) between the transmitting and receiving devices and the adopted propagation model.

The path loss \(\text{PL}(d_i)\) is modeled for complex \ac{NLOS} propagation environments using the \ac{LDPL} model~\cite{rappaport1996wireless}:
\begin{equation}\label{eq:LDPL}
\text{PL}_{\text{LDPL}}(d) = PL_0 + 10 \gamma \log_{10}\left(\frac{d}{d_0}\right) + X_\sigma,    
\end{equation} 
where \(PL_0\) is the reference path loss at a reference distance \(d_0 = 1\) m, \(\gamma\) is the path loss exponent, and \(X_\sigma\) is the shadowing effect modeled as a Gaussian random variable. 

\subsubsection{Noise Floor Modeling}

The noise floor (\(\eta_i^t\)) at device \( i \) is the combination of baseline environmental noise and jammer interference, determined using the relevant path loss model. It is expressed in dBm as:
\[
\eta_i^t = 10 \cdot \log_{10}(N_{\text{ambient}} + P_{\text{jam},i}),
\]
where \(N_{\text{ambient}}\) is the ambient noise level, and \(P_{\text{jam},i}\) is the jammer \ac{RSSI} at device \( i \), both represented in mW. The ambient noise level is assumed to be -100 dBm. The jammer's \ac{RSSI} is computed as:
\[
P_{\text{jam},i} = P_t^{\text{jam}} + G_t^{\text{jam}} + G_r - \text{PL}(d_{\text{jam},i}),
\]
where \(P_t^{\text{jam}}\) is the jammer’s transmit power, \(G_t^{\text{jam}}\) is its transmitting antenna gain, and \(d_{\text{jam},i}\) is the distance from the jammer at location \( \mathbf{x}_j \) to device \( i \).

\subsubsection{Data Generation}
We generate 2D (static) and 3D (dynamic) datasets to evaluate localization algorithms. For the static, we create 25,000 samples per topology per jammer placement strategy and 1000 instances for the dynamic dataset, both utilizing LDPL to model signal propagation. Urban and shadowed urban areas are modeled, with path loss exponents ranging from 2.7 to 3.5 and 3.0 to 5.0~\cite{aldosari2019jtx}, respectively, and moderate to strong shadowing effects between 2 to 6 dB~\cite{tedeschi2022pwrmodjam}, representing physical phenomena which impact the accuracy of signal-based localization techniques~\cite{aldosari2019jtx}. In the static dataset, nodes are placed within the geographic area \(\mathcal{A} = \{ (x, y) \in \mathbb{R}^2 \mid 0 \leq x, y \leq 1500 \}\), with a communication range of 200m. The nodes are arranged in circular, triangular, rectangular, and uniformly random topologies, where the circular and triangular topologies feature a perimeter-bounded node distribution, while a surface-covering node distribution characterizes the rectangular and random topologies. The number of nodes is determined using a beta-distributed scaling factor within a calculated minimum and maximum range, allowing for variability in network densities to simulate different levels of congestion or sparsity. The dataset captures scenarios of both fully jammed or partially jammed networks, ensuring at least one node records a noise floor above -80 dBm, and at least three nodes are jammed to distinguish jamming effects from natural signal fading.  For the dynamic dataset, the geographical area of the trajectory is \(\mathcal{A} = \{ (x, y, z) \in \mathbb{R}^3 \mid 0 \leq x, y \leq 500, \, 0 \leq z \leq 50 \}\), though device motion may extend beyond these bounds during simulation. Further information regarding dataset statistics can be found in~\ref{tab:dataset_statistics}. For reproducibility, we share the data at \textbf{\textit{\url{https://www.kaggle.com/datasets/daniaherzalla/network-jamming-simulation-dataset}}}.

\begin{table}
    \centering
    \scriptsize
    \caption{Datasets statistics reported for number of nodes per graph (Nodes), noise floor (\( \eta \)), jammer transmit power (\(P_t^{\text{jam}}\)), shadowing (\(X_\sigma\)), path loss exponent (\(\gamma\)), and distance of graph centroid to jammer (\(D_{jam}\)).}\label{tab:dataset_statistics}
    \begin{tabularx}{\columnwidth}{@{}>{\hsize=0.35\hsize}X>{\hsize=0.85\hsize}X>{\hsize=0.6\hsize}C>{\hsize=0.6\hsize}C} 
        \toprule
        Data & Feature & Mean & Min/Max \\
        \midrule
        \multirow{5}{*}{\rotatebox[origin=c]{90}{Static}} 
        & Nodes & 33.6\sd{19.1} & 3.0/122.0 \\
        & \( \eta \) & -63.9\sd{13.3} & -100.0/72.4 \\
        & \(P_t^{\text{jam}}\) & 51.1\sd{6.3} & 20.0/59.0 \\
        & \(X_\sigma\) & 4.2\sd{1.2} & 2.0/6.0 \\
        & \(\gamma\) & 3.1\sd{0.3} & 2.7/5.0 \\
        & \(D_{jam}\) & 674.6\sd{392.6} & 0.1/2117.4 \\
        \midrule
        \multirow{5}{*}{\rotatebox[origin=c]{90}{Dynamic}} 
        & Nodes & 5809.0\sd{5527.7} & 1112.0/37144.0 \\
        & \( \eta \) & -56.6\sd{18.6} & -80.0/19.9 \\
        & \(P_t^{\text{jam}}\) & 39.3\sd{11.9} & 20.0/60.0 \\
        & \(X_\sigma\) & 4.0\sd{1.1} & 2.0/6.0 \\
        & \(\gamma\) & 3.1\sd{0.2} & 2.7/3.5 \\
        & \(D_{jam}\) & 295.2\sd{349.5} & 3.0/2184.7 \\
        \bottomrule
    \end{tabularx} 
\end{table}

\end{document}